\documentclass[preprintnumbers,amsmath,amssymb]{revtex4}
\usepackage{graphicx}
\usepackage{makeidx}
\usepackage{amssymb}
\usepackage{subfigure}
\usepackage{chemarr}
\usepackage{bm}

\def\epsilon{\varepsilon}

\def\i{\indent}

\def\beqr{\begin{eqnarray}}
\def\eqnr{\end{eqnarray}}
\def\beq{\begin{equation}}
\def\bc{\begin{center}}
\def\ec{\end{center}}
\def\eqn{\end{equation}}

\begin{document}
\title{Analysis of Remote Synchronization in Complex Networks}
\author{Lucia Valentina Gambuzza$^1$, Alessio Cardillo$^{2,3}$, Alessandro Fiasconaro$^{2,4}$, Luigi Fortuna$^1$, Jesus G{\'o}mez-Garde\~{n}es$^{2,3}$, Mattia Frasca$^1$}
\affiliation{$^1$ Dipartimento
di Ingegneria Elettrica Elettronica e Informatica, Universit\`a degli Studi di Catania, viale A. Doria 6,
95125 Catania, Italy\\
$^2$ Departamento de F\'{\i}sica de la Materia Condensada,
University of Zaragoza, Zaragoza 50009, Spain\\
$^3$ Institute for Biocomputation and Physics of Complex
Systems (BIFI), University of Zaragoza, Zaragoza 50018, Spain\\
$^4$ Instituto de Ciencia de Materiales de Arag\'on, CSIC--University of Zaragoza, Zaragoza 50009, Spain}

\begin{abstract}
A novel regime of synchronization, called remote synchronization, where the peripheral nodes form a phase synchronized cluster not including the hub, was recently observed in star motifs \cite{RSpre}. We show the existence of a more general dynamical state of remote synchronization in arbitrary networks of coupled oscillators. This state is characterized by the synchronization of pairs of nodes that are not directly connected via a physical link or any sequence of synchronized nodes. This phenomenon is almost negligible in networks of phase oscillators as its underlying mechanism is the modulation of the amplitude of those intermediary nodes between the remotely synchronized units. Our findings thus show the ubiquity and robustness of these states and bridge the gap from their recent observation in simple toy graphs to complex networks.
\end{abstract}


\maketitle

\textbf{In this work we show a novel synchronization state in networks of coupled oscillators. This state, called Remote Synchronization, is characterized by the synchronization of pairs of nodes that are not directly connected via a physical link or any sequence of synchronized nodes. Moreover, remote synchronization is manifested when considering oscillators having amplitude and phase as dynamical variables, in contrast to the usual setting in which phase oscillators are considered, as its underlying mechanism is the modulation of the amplitude of those intermediary nodes allowing the exchange of information between remotely synchronized units. Although some previous observations of such phenomenon were made in simple star-like graphs, here we show its ubiquity in the general framework of complex networks. To this end we analyze its existence as a robust dynamical state that appears before global synchronization shows up. Our findings thus open the door for experimental observations of this novel state in which the existence of a synchronized pair cannot be associated to a given physical interaction through a single link of the network. In addition, our results highlight the important difference between the real (i.e. associated to physical links) and the functional (i.e. emerging from synchronization) connectivity of a network.}

\section{Introduction}

Synchronization constitutes one of the most paradigmatic examples of emergence of collective behavior in natural, social and man-made systems \cite{strogatzrev,piko,boccbook}. Its ubiquity relies on the general framework in which it occurs: the interaction between two or more nonidentical dynamical units that, as a consequence, adjust a given property of their motion. As coupling between units increases, synchronization shows up as a collective state in which the units behave in a coordinated way. Synchronization phenomena span across many life scales, ranging from the development of cognitive tasks in neural systems \cite{bullmore} to the onset of social consensus in human societies \cite{latora_opinion}.

The ubiquity of synchronization in real systems together with the recent discovery \cite{strogatz,physrepbocca,doro08,wu12,ach12} of their real architecture of interactions has motivated its study when units are embedded in a complex network~\cite{PRsync}. In this way, each unit is represented as a node of a network while it only interacts with those adjacent units, {\em i.e.} those directly coupled via an edge. In the last decade many studies have unveiled the impact that diverse interaction topologies have on the onset of synchrony~\cite{Pacheco,Arenas,gardenes,lodato,explosive} and its stability~\cite{nishikawa,chavez,zhou,delosrios}. In addition, related issues such as that of adaptive networks, in which the interaction pattern changes according to the degree of synchronization of the system, have also attracted the attention of the community~\cite{ZhouPRL06,OttPRL08,AokiPRL09,BoccPRL11}.

The former studies mainly rely on the study of coupled phase oscillators, such as the Kuramoto model \cite{kuramoto,acebron}, which produces globally synchronized systems as a result of the direct interaction of pairs of adjacent units. However, it has been recently found \cite{RSpre} that, for more general oscillator models (in which both amplitude and phase are dynamical variables) such as the Stuart-Landau (SL) model \cite{piko}, two oscillators, that are not directly linked but are both connected to a third unit, can become synchronized even if the third oscillator does not synchronize with them. This novel phenomenon, termed \emph{remote synchronization}, relies on the modulation of the amplitude parameter of an intermediary node allowing the passage of information between two of its neighbors for their synchronization, even when the former is not synchronized with them. Thus, this tunnel-like mechanism is out of reach in ensembles of phase oscillators. Although the term remote synchronization has been used in quite different contexts, as for example in computer science where it refers to synchronization of two or more files located in two, remotely connected, computers or in some synchronization schemes for dynamical systems \cite{cinese}, to emphasize the remote location of the receiver with respect to the transmitter, we will use it to refer to the novel form of synchronization as reported in \cite{RSpre}.

Remote synchronization has been found to occur in very specific and simple topologies such as star-like networks in which the central node has a natural frequency different from that of the leaves. Within this particular setting it was numerically and experimentally shown \cite{RSpre} that leaves become mutually synchronized without the need of the synchronization of the central node. In this paper, we aim at showing that remote synchronization is not limited to the particular configuration of a star-like motif or a tight specification of the node frequencies. To this end, we introduce a general procedure for detecting remote synchronization in arbitrary networks and then discuss the results of our analysis on arbitrary complex networks.

\section{Measures of remote synchronization}

In \cite{RSpre}, where star motifs were dealt with, remote synchronization is detected by observing that for intermediate values of the coupling coefficient the synchronization level among the leaves (measured with the so called \emph{indirect} Kuramoto parameter) is higher than that between the hub and the leaves (measured with the so called \emph{direct} Kuramoto parameter). We note that such measures are not applicable to the general case of arbitrary topologies, since they are based on an a priori analysis of the network structure which allows one to establish which nodes can remotely synchronize. Therefore, in this paper we first introduce a general procedure for detecting remote synchronization in arbitrary networks and then show ubiquity and robustness of remote synchronization in the general case of complex networks.

To this end, we consider a network of $N$  coupled Stuart-Landau oscillators \cite{piko}. Each node $i$ is characterized by two variables, $(x_i, y_i)^T$, whose dynamical evolution is as follows:
\begin{equation}
\begin{array}{cc}
\left ( \begin{array}{r} \dot{x}_i \\ \dot{y}_i \end{array} \right )=\left ( \begin{array}{cc} \alpha - x_i^2 - y_i^2 & -\omega_i \\ \omega_i & \alpha -x_i^2-y_i^2 \end{array} \right )\left ( \begin{array}{l} x_i \\ y_i \end{array} \right )
+\frac{\lambda}{k_i} \sum_{j=1}^N a_{ij}
\left [
\left ( \begin{array}{l} x_j \\ y_j \end{array} \right )
-\left ( \begin{array}{l} x_i \\ y_i \end{array} \right )
\right ]
\;,
\end{array}
\label{eq:networkeqs}
\end{equation}
where $\sqrt{\alpha}$ and $\omega_i$ are respectively the amplitude and the (natural) frequency of oscillator $i$ when uncoupled. The second term on the right accounts for the coupling of the dynamics of node $i$ with its $k_i$  neighbors. The strength of the coupling is controlled by $\lambda$ ($\lambda=0$ in the uncoupled limit) while ${\cal A}=\{a_{ij}\}$ represents the adjacency matrix of the network defined as: {\em (i)} for $i\neq j$, $a_{ij}=1$ when nodes $i$ and $j$ are connected while $a_{ij}=0$ otherwise, and {\em (ii)} $a_{ii}=0$.

\begin{figure}
\centering
\subfigure[]{\includegraphics[height=.18\textheight]{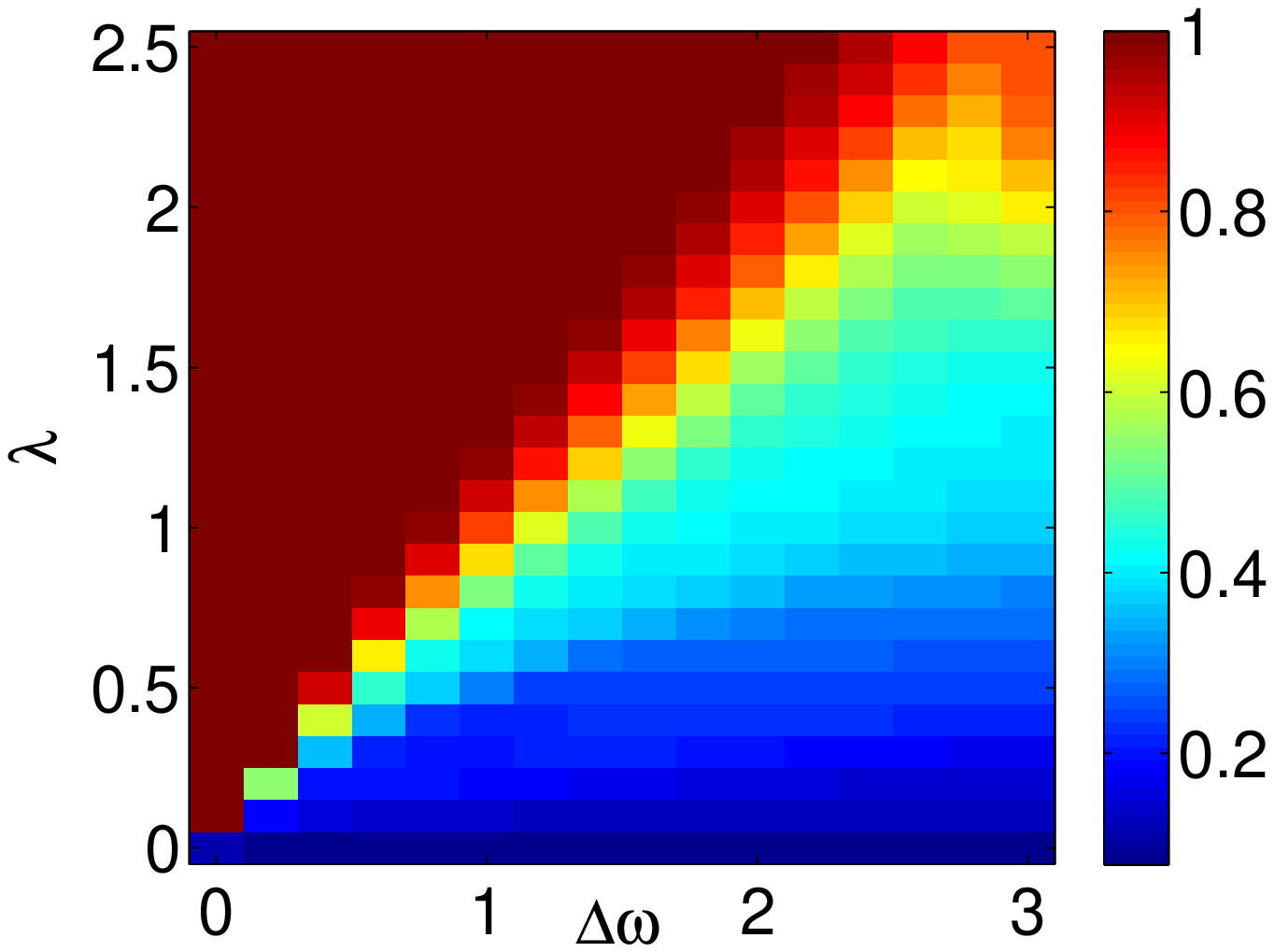}\label{fig:diagrammi2D_A}}
\subfigure[]{\includegraphics[height=.18\textheight]{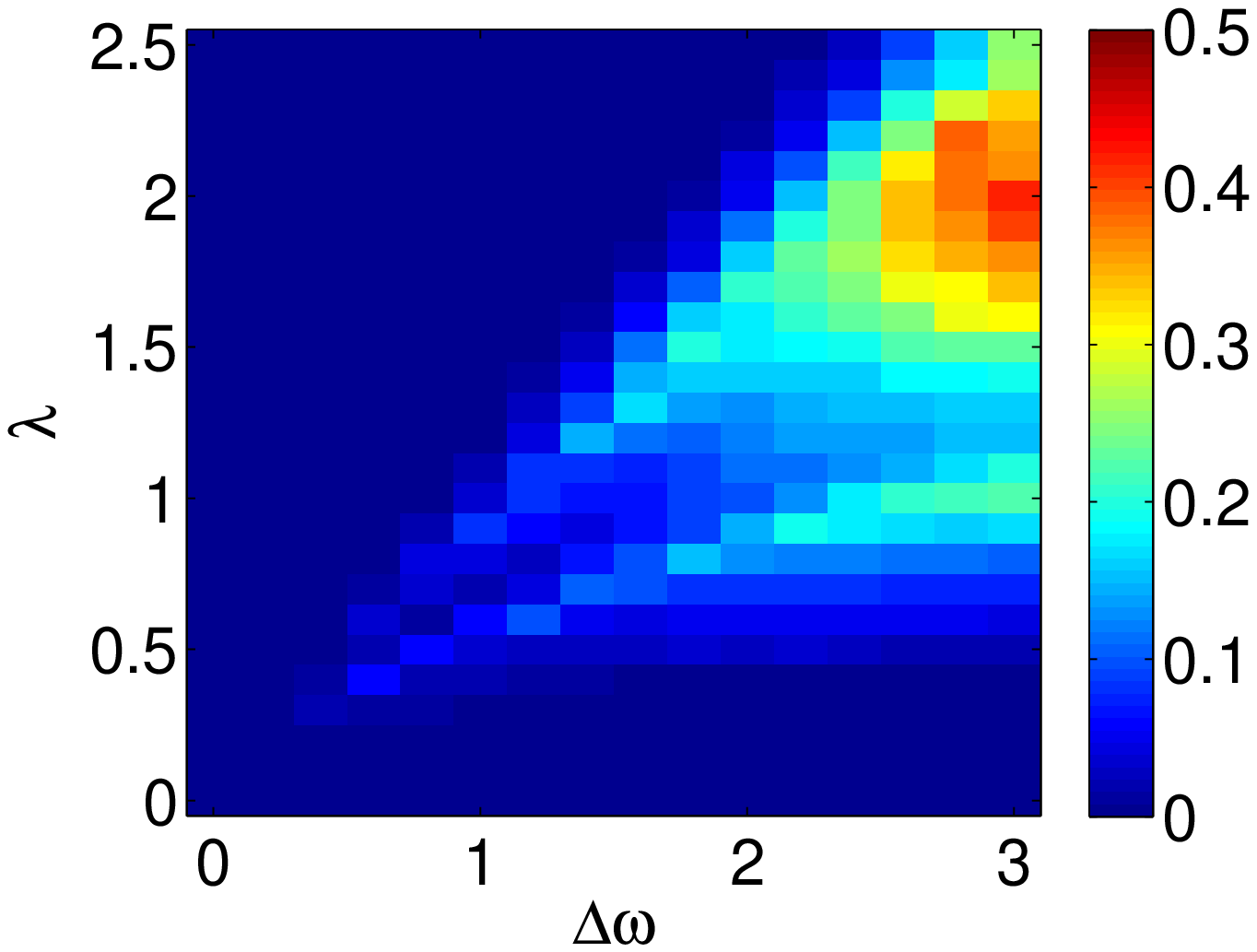}\label{fig:diagrammi2D_B}}\\
\subfigure[]{\includegraphics[height=.18\textheight]{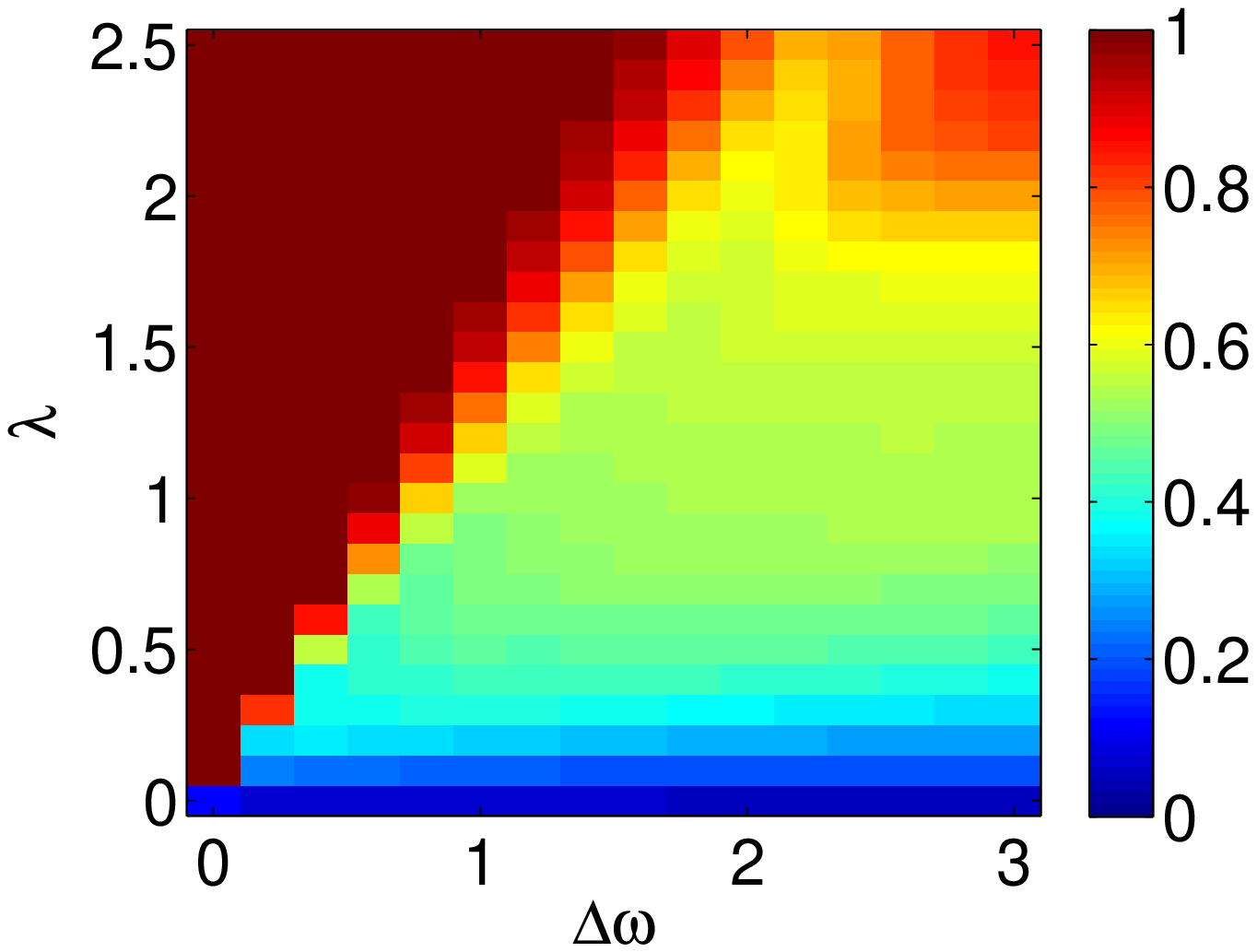}\label{fig:diagrammi2D_C}}
\subfigure[]{\includegraphics[height=.18\textheight]{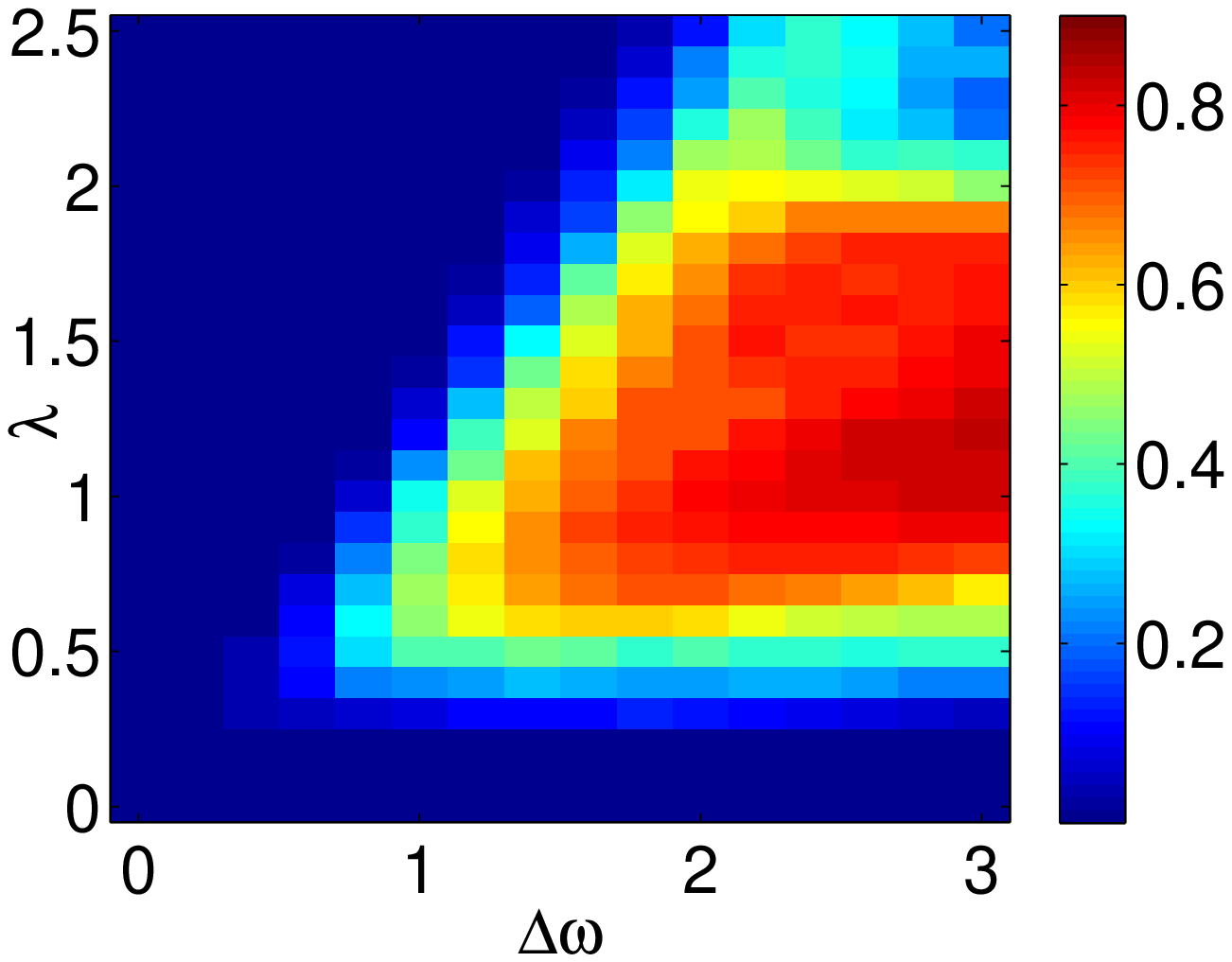}\label{fig:diagrammi2D_D}}
\caption{(color online). Evolution of the degree of global synchronization $r$ [panels (a) and (c)] and the number of remotely synchronized nodes $n_{RS}$ [panels (b) and (d)] for SF (upper panels) and ER (bottom panels) networks as a function of the coupling strength $\lambda$ and the frequency mismatch $\Delta \omega$. In both cases the networks have $N=100$ and $<k>=2$. The other relevant parameters are fixed to $\alpha=1$, $\omega_l=1$. Remote synchronization (high values of $n_{RS}$) is found for strong frequency mismatch $\Delta \omega$ and moderate coupling $\lambda$, while, for low values of the coupling parameter, nodes cannot synchronize ($r$ and $n_{RS}$ have low values), and, for large values of $\lambda$, the network is fully synchronized ($r\simeq 1$).
\label{fig:diagrammi2D}}
\end{figure}

To study the synchronization properties of system (\ref{eq:networkeqs}) we work with the phase variable of each oscillator, defined as
$\theta_i=\tan^{-1}{(y_i/x_i)}$. Then we can measure the degree of synchronization of any (connected or not) pair of oscillators by means of the time averaged order parameter:
\begin{equation}
r_{ij}=|\langle e^{\iota[\theta_i(t)-\theta_j(t)]}\rangle_t|\;,
\end{equation}
where $\langle \cdot \rangle_t$ means an average over a large enough time interval and $\iota=\sqrt{-1}$. We will consider two nodes as synchronized when $r_{ij}>\delta$, where $\delta$ is a constant threshold that we fix to $\delta=0.8$. Nonetheless, we checked that the results presented are robust as other values of $\delta$ yield qualitatively the same outcomes.

Once two nodes $i$ and $j$ are classified as mutually synchronized we label their relationship according to the following three situations: {\em (i)} $i$ and $j$ are directly connected ($a_{ij}=a_{ji}=1$), {\em (ii)} there is a path of mutually synchronized nodes between them, and {\em (iii)} neither of the former two situations hold. While the first two cases are similar, as both are examples of synchronization through \emph{physical links}, the third case is analogous to the observed remote synchronization in a star-like network, but in the more general context of a complex network. Thus, we define that two nodes $i$ and $j$ are remotely synchronized (RS) when they are synchronized ($r_{ij}>\delta$) and they are not connected by either a direct link or a path of synchronized nodes.

To quantify systematically the extent of remote synchronization we count the number of RS nodes, defined as the number $N_{RS}$ of nodes that appear RS with at least another node in the network. This allows us to introduce the following order parameter:
$n_{RS}=N_{RS}/N$, representing the normalized number of RS nodes with respect to the total number of nodes $N$.
%
Finally, to quantify the importance that remote synchronization has on the dynamics of the system
we also measure the global level of synchronization through the usual Kuramoto-like order parameter: 
\begin{equation}
\label{eq:kuramotoparameter}
r=\frac{1}{N^2}\sum_{i,j=1}^N{r_{ij}}\;.
\end{equation}
Note that $r$
takes into account the contribution of both synchronized ($r_{ij}>\delta$) and not synchronized ($r_{ij}\leq \delta$) nodes.

\section{Results}
As two well-known paradigmatic network topologies we have analyzed both Erd\H{o}s-R\'enyi (ER) and Scale-free (SF) graphs. The former type of networks is characterized by a Poisson distribution $P(k)$ for the probability of finding a node with $k$ contacts while SF graphs show a power-law distribution, $P(k)\sim k^{-\gamma}$. Thus, while in ER graphs most of the nodes are close to the mean connectivity $\langle k\rangle$, SF networks display a large heterogeneity in the number of contacts per node as revealed from the existence of hubs having $k_i\gg \langle k\rangle$. For their construction we have made use of the model introduced in \cite{GGM} that allows one to control the mean connectivity of both networks in order to be exactly the same. In the networks reported in this paper the size and mean connectivity are fixed to $N=100$ and $\langle k \rangle=2$ respectively. The SF networks generated with this model have $\gamma=3$. Finally, in order to stay close to the framework used in \cite{RSpre} we have considered a bimodal distribution for the natural frequencies of the oscillators so that nodes with high degree (those analogous to the central nodes in a star graph) present a larger frequency, $\omega_h$, than that, $\omega_l$, of less connected (the ones playing the role of leaves in the star topology). In particular, we labeled as hubs those nodes having $k_i>k^*$ \cite{footnote} and assigned them $\omega_i=\omega_h + \xi_i\omega_h$  while, for the rest of nodes $\omega_i=\omega_l+\xi_i\omega_l$. In the former expressions $\xi_i$ is a random variable uniformly distributed between -0.025 and 0.025.

In Fig.~\ref{fig:diagrammi2D} we show the emergence of remote synchronization as a function of the two relevant parameters: the coupling strength $\lambda$ and the frequency mismatch of the network hubs $\Delta \omega=\omega_h-\omega_l$. In particular, we report the behavior of the global synchronization, $r$, [panels \ref{fig:diagrammi2D_A} and \ref{fig:diagrammi2D_C}] and the fraction of RS nodes, $n_{RS}$, [panels \ref{fig:diagrammi2D_B} and \ref{fig:diagrammi2D_D}] for SF (top) and ER (bottom) networks. The results are averaged over $50$ different network instances and, for each network we average the results over $10 $ different realizations of the distribution of natural frequencies.

We find that remote synchronization occurs in both types of networks in a region of parameters characterized by a strong frequency mismatch $\Delta \omega$ and moderate coupling $\lambda$. In fact, for low values of the coupling parameter,
nodes cannot synchronize (either in a direct or remote way) as observed from the low values of $r$ and $n_{RS}$.
On the contrary, for large values of $\lambda$ the network is fully synchronized ($r\simeq 1$) and, accordingly, $n_{RS}$ assumes values close to zero since all the nodes are mutually synchronized with their neighbors. As panels~\ref{fig:diagrammi2D_A} and~\ref{fig:diagrammi2D_C} reveal, the onset of full synchronization requires greater values of the coupling as the frequency mismatch increases. In fact, a large frequency mismatch together with values of coupling under the threshold for complete synchronization favors the onset of remote synchronization, as observed from the behavior of $n_{RS}$ in panels~\ref{fig:diagrammi2D_B} and \ref{fig:diagrammi2D_D}.

Compared to SF networks, the values of $n_{RS}$ in ER networks are greater, thus indicating that remote synchronization in ER networks involves a larger number of nodes. Moreover, in ER networks the onset of remote synchronization occurs for lower values of $\lambda$. ER and SF networks also show qualitative differences in the appearance of remote synchronization: by keeping fixed $\Delta \omega$ and increasing the value of $\lambda$, we find that $n_{RS}$ in SF networks show two peaks, while for ER networks it shows a rise-and-fall behavior.

\begin{figure}
\centering
 \subfigure[]{\includegraphics[height=.2\textheight]{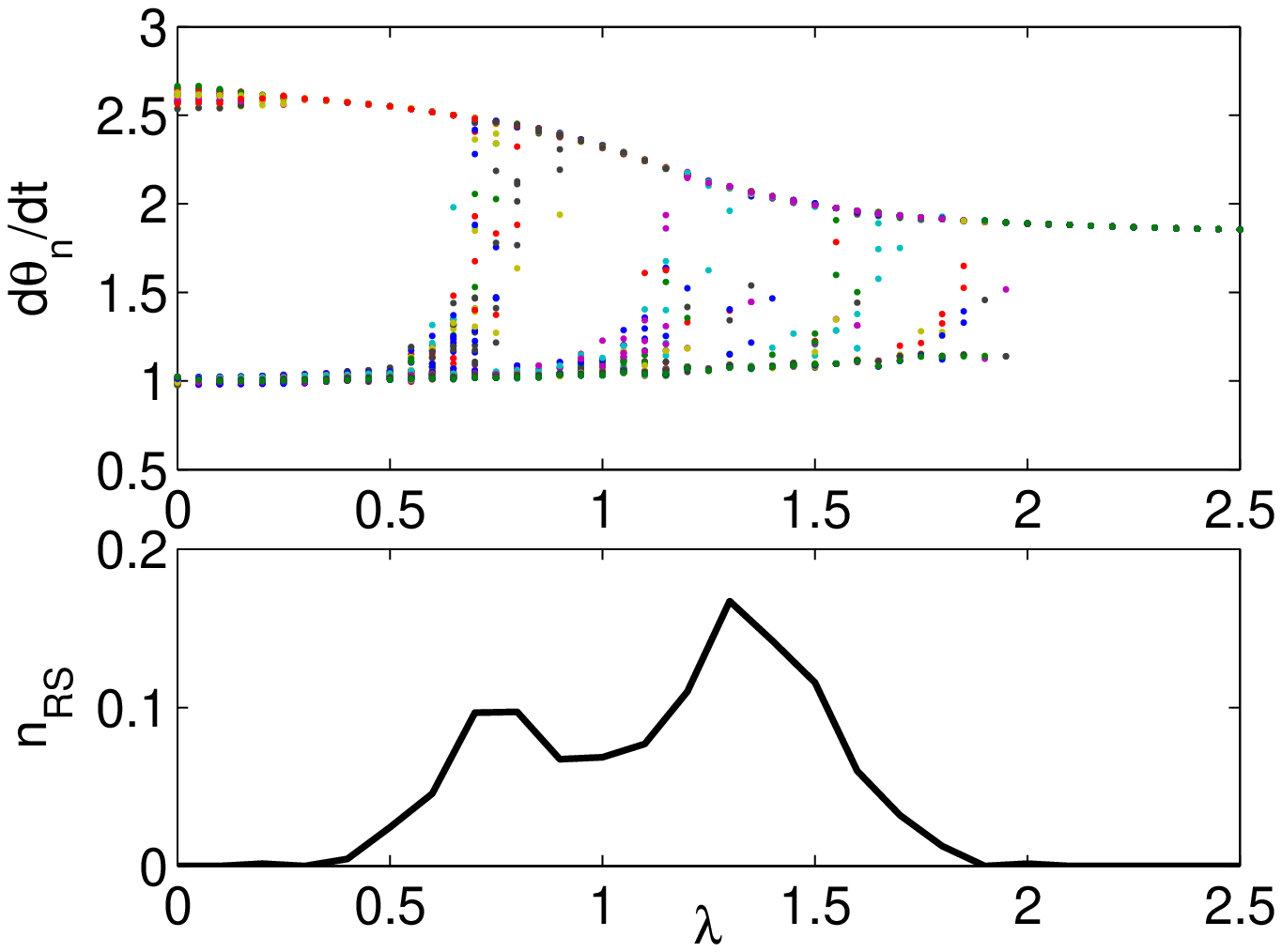}\label{fig:freqbehaviorA}}
  \subfigure[]{\includegraphics[height=.2\textheight]{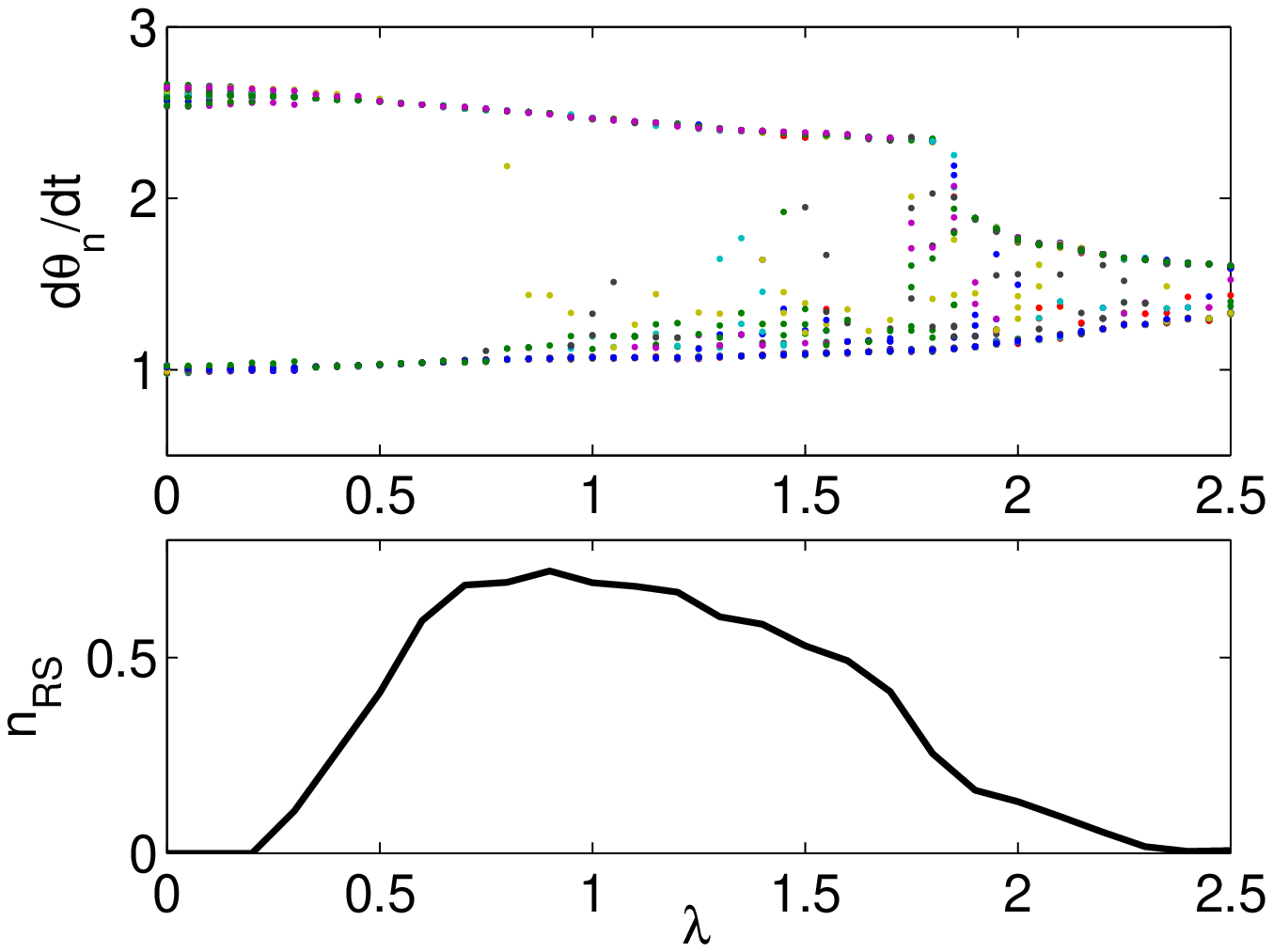}\label{fig:freqbehaviorB}}
\caption{(color online). Evolution of the average oscillation frequency of each oscillator and $n_{RS}$ as a function of $\lambda$ for SF (a) and ER (b) networks. The mismatch of natural frequencies is $\Delta \omega=1.5$ while the rest of parameters are the same as in Fig.~\ref{fig:diagrammi2D}.
The average oscillation frequencies, which for $\lambda=0$ start from a bimodal distribution as dictated by the configuration for the natural
frequencies, as $\lambda$ is increased tend towards a common value, characterizing full synchronization.
The strong reorganization of the frequencies (characterized by a spread of the oscillation frequencies between the two extreme values) corresponds to the values of coupling for which $n_{RS}$ is peaked.
\label{fig:freqbehavior}}
\end{figure}

In both (ER and SF) cases remote synchronization appears as an intermediary state before full synchronization is achieved. However, from the analysis of panels~\ref{fig:diagrammi2D_A} and \ref{fig:diagrammi2D_C} one observes that the behavior of $r$ {\em vs.} $\lambda$ for a fixed value of $\Delta \omega$ is qualitatively different in SF and ER networks. In particular, in ER networks (panel \ref{fig:diagrammi2D_C}) a large plateau around $r\simeq 0.5$ is set in the region where remote synchronization shows up. In this region, the increase of $\lambda$ does not contribute to the overall synchronization level, but to a redistribution of the average oscillation frequencies of the network nodes.

This is evident in Fig.~\ref{fig:freqbehavior}, where the average values (over the simulation time $T$) of the instantaneous frequency of each oscillator are reported along with the parameter $n_{RS}$. The results are obtained by increasing $\lambda$ adiabatically from $\lambda=0$ so that the system starts from a bimodal distribution as dictated by the configuration for the natural oscillations. As $\lambda$ increases, the gap between the two main frequency values of the bimodal distribution decreases until the network reaches full synchronization and the nodes oscillate at a common frequency. The readjustment of frequencies reveals that, for some values of the coupling, the system undergoes a strong reorganization, as shown by the spread of the oscillation frequencies between the two extreme values. This readjustment coincides with the peaks displayed by $n_{RS}$ in both SF and ER networks. However, the readjustment seems to occur faster in SF networks for which the plateau of $r$ is not observed.

Now we illustrate the role of parameter $\alpha$. To this end, we consider a general graph and show that
for $\alpha>>1$ the SL model transforms into a network of Kuramoto oscillators, so that the amplitude of the oscillators become decoupled and stationary. We consider Eqs. (\ref{eq:networkeqs}) in polar coordinates:

\begin{equation}
\label{eq:StuartLandauOscillatorLambdasuKNodePolar}
\left.%
\begin{array}{l} \dot{\rho}_i=\alpha \rho_i -\rho_i^3 + \frac{\lambda}{k_i} \sum_{j=1}^N a_{ij}(\rho_j \cos(\theta_j-\theta_i)-\rho_i) \\
\dot{\theta}_i=\omega_i+\frac{\lambda}{k_i} \sum_{j=1}^N\frac{\rho_j}{\rho_i}a_{ij}\sin(\theta_j-\theta_i)
\end{array}
\right.
\end{equation}

\noindent where $\rho_i e^{\iota\theta_i}=x_i+\iota y_i$. Defining $R_{i}=\frac{\rho_{i}}{\sqrt\alpha}$, where $\sqrt\alpha$ is the value of the amplitude at the equilibrium, Eqs.(\ref{eq:StuartLandauOscillatorLambdasuKNodePolar}) can be rewritten as follows:

\begin{equation}
\label{eq:StuartLandauOscillatorLambdasuKNodePolar2}
\left.%
\begin{array}{l} \dot{R}_i=\alpha R_i -\alpha R_i^3 + \frac{\lambda}{k_i} \sum_{j=1}^N a_{ij}(R_j \cos(\theta_j-\theta_i)-R_i) \\
\dot{\theta}_i=\omega_i+\frac{\lambda}{k_i} \sum_{j=1}^N\frac{R_j}{R_i}a_{ij}\sin(\theta_j-\theta_i)
\end{array}
\right.
\end{equation}

In the first equation we can rescale time according to $dT=\alpha dt$
(while the second equation remains unchanged).

\begin{equation}
\label{eq:StuartLandauOscillatorLambdasuKNodePolar4}
\left.%
\begin{array}{l} \frac{{dR}_i}{dT}=R_i - R_i^3 + \frac{\lambda}{\alpha k_i} \sum_{j=1}^N a_{ij}(R_j \cos(\theta_j-\theta_i)-R_i) \\
\dot{\theta}_i=\omega_i+\frac{\lambda}{k_i} \sum_{j=1}^N\frac{R_j}{R_i}a_{ij}\sin(\theta_j-\theta_i)
\end{array}
\right.
\end{equation}

Now as $\alpha\rightarrow\infty$ the coupling term in the amplitude
equation vanishes, and from the analysis of the first equation we derive
that $R_i\rightarrow 1$ for all $i$ (in fact $R_i=1$ is the only
equilibrium and the dynamics evolve very fast as $dT=\alpha dt$ and
$\alpha$ is large). In the second equation $R_i\rightarrow 1$ leads to
$\frac{R_i}{R_j}=1$ and thus the second equation becomes:

\begin{equation}
\label{eq:StuartLandauOscillatorLambdasuKCentralNodePolar5}
\dot{\theta}_i=\omega_i+\frac{\lambda}{k_i} \sum_{j=1}^N a_{ij}\sin(\theta_j-\theta_i)
\end{equation}

\begin{figure}
\centering
\subfigure[]{\includegraphics[height=.2\textheight]{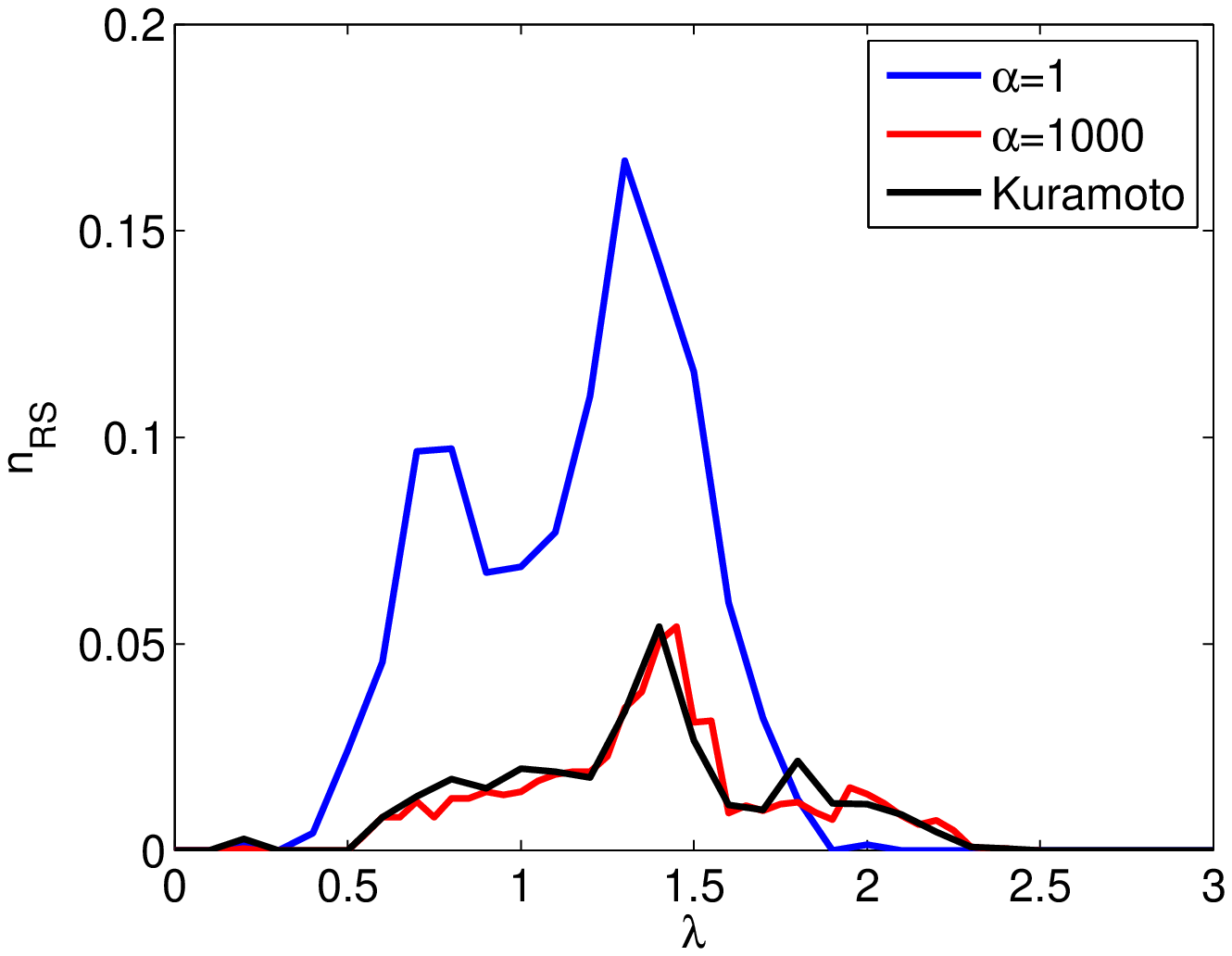}\label{fig:nodi_SF}}
\subfigure[]{\includegraphics[height=.2\textheight]{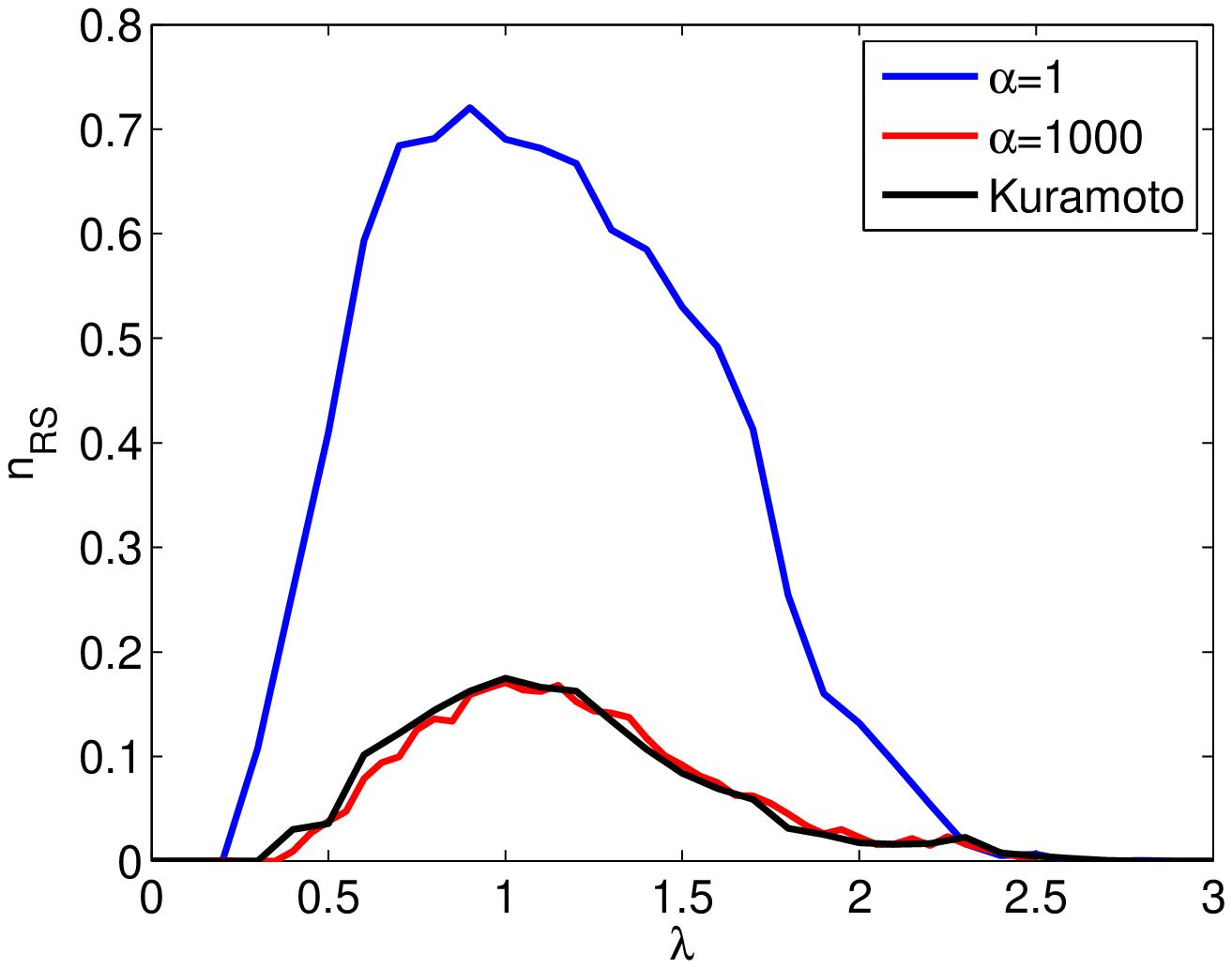}\label{fig:nodi_ER}}\\
\subfigure[]{\includegraphics[height=.2\textheight]{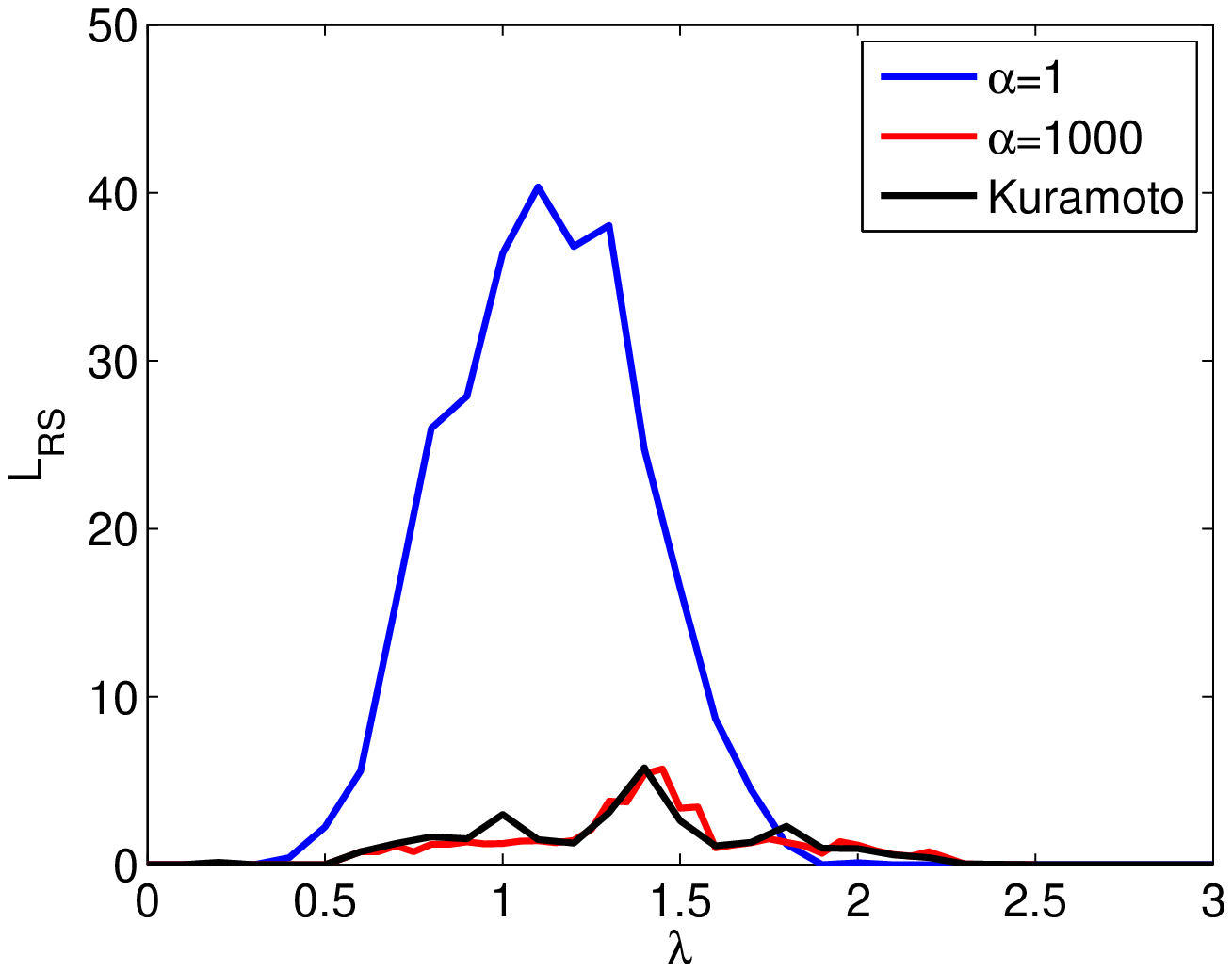}\label{fig:link_SF_nonnormalizzati}}
\subfigure[]{\includegraphics[height=.2\textheight]{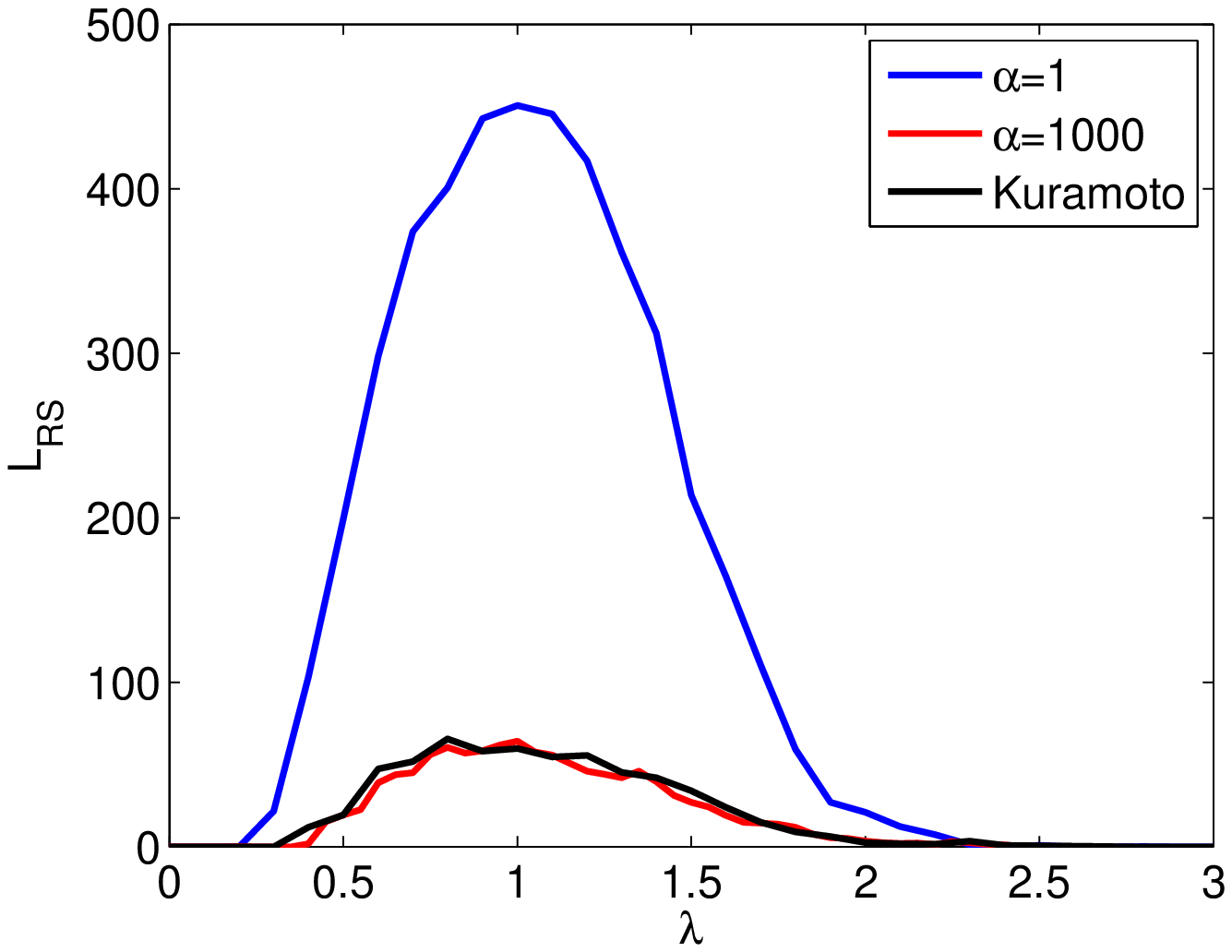}\label{fig:link_ER_nonnormalizzati}}
\caption{(color online). Comparison of $n_{RS}$ (a-b) and $L_{RS}$ (c-d) in a SF (a,c) and ER (b,d) network of Stuart-Landau oscillators with $\alpha=1$ or $\alpha=1000$ and a network of Kuramoto purely phase oscillators for
$\Delta \omega=2.6$. \label{fig:confrontoKuramoto}}
\end{figure}

Therefore, as $\alpha\rightarrow\infty$, we recover the model of Kuramoto purely phase oscillators coupled into a network. In this limit, we observe that the amplitude equation plays no role. In this case, the level of RS is very low, as for instance reported in Fig.~\ref{fig:confrontoKuramoto}, where a network of Stuart-Landau oscillators with $\alpha=1$ is compared with a network of Stuart-Landau oscillators with $\alpha=1000$ and with a network of Kuramoto purely phase oscillators. We note that for $\alpha=1000$ the network of Stuart-Landau oscillators is already a good approximation of the network of Kuramoto purely phase oscillators. In both the two examples of networks (SF and ER), for Kuramoto oscillators $n_{RS}$ (Fig.~\ref{fig:nodi_SF}-\ref{fig:nodi_ER}) is lower than in Stuart-Landau oscillators (with $\alpha=1$). The lower level of RS in Kuramoto oscillators is more evident when the number of RS links, labeled as $L_{RS}$, is examined as in Fig.~\ref{fig:link_SF_nonnormalizzati}-\ref{fig:link_ER_nonnormalizzati}, which shows how the number of RS links is decreased by an order of magnitude with respect to the case of Stuart-Landau oscillators (with $\alpha=1$). This suggests that amplitude modulation is the main mechanism underlying RS (this was also shown with other arguments in \cite{RSpre} for star-like networks).

\begin{figure}
\centering
 \subfigure[]{\includegraphics[height=.2\textheight]{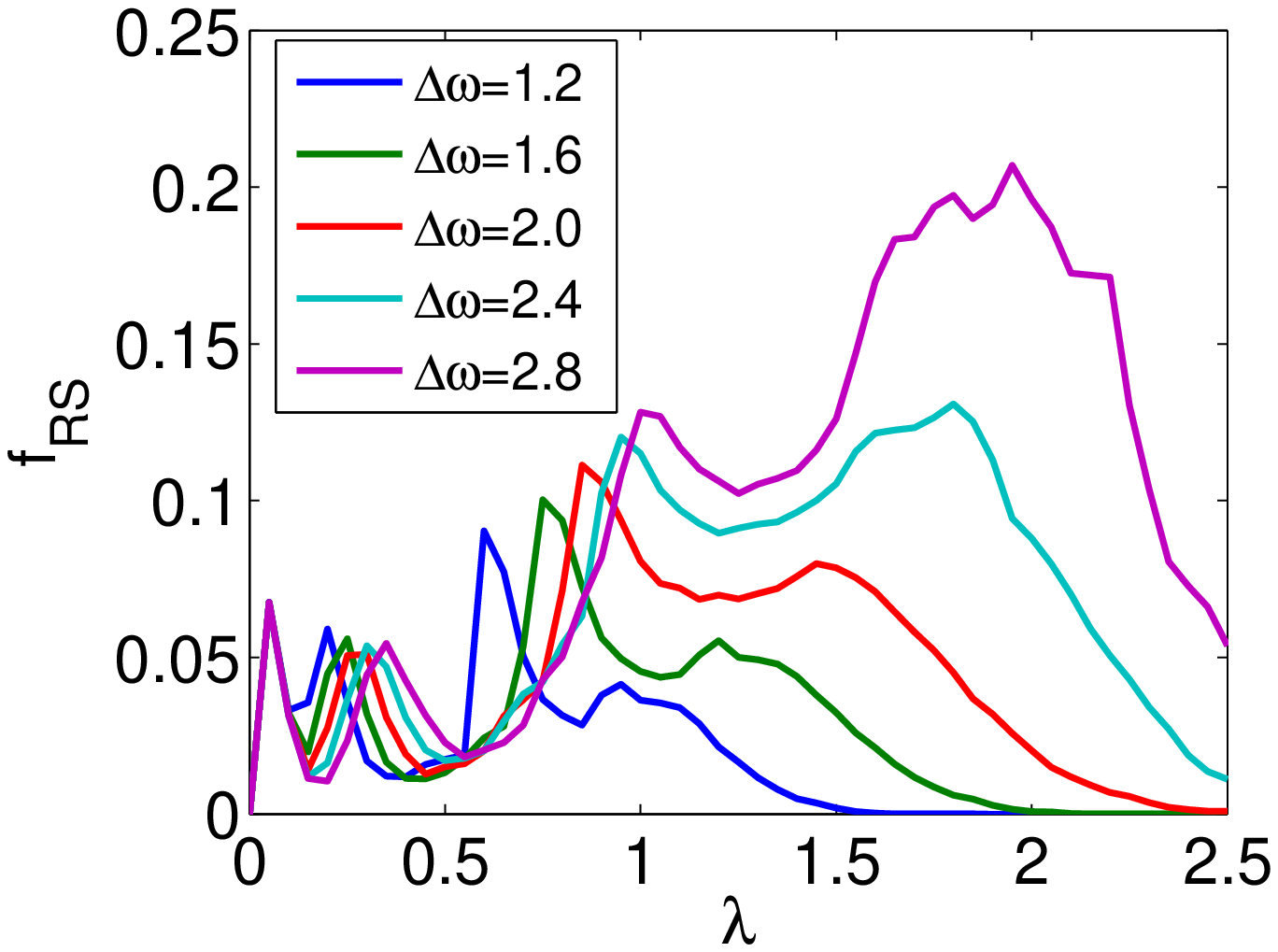}}
 \subfigure[]{\includegraphics[height=.2\textheight]{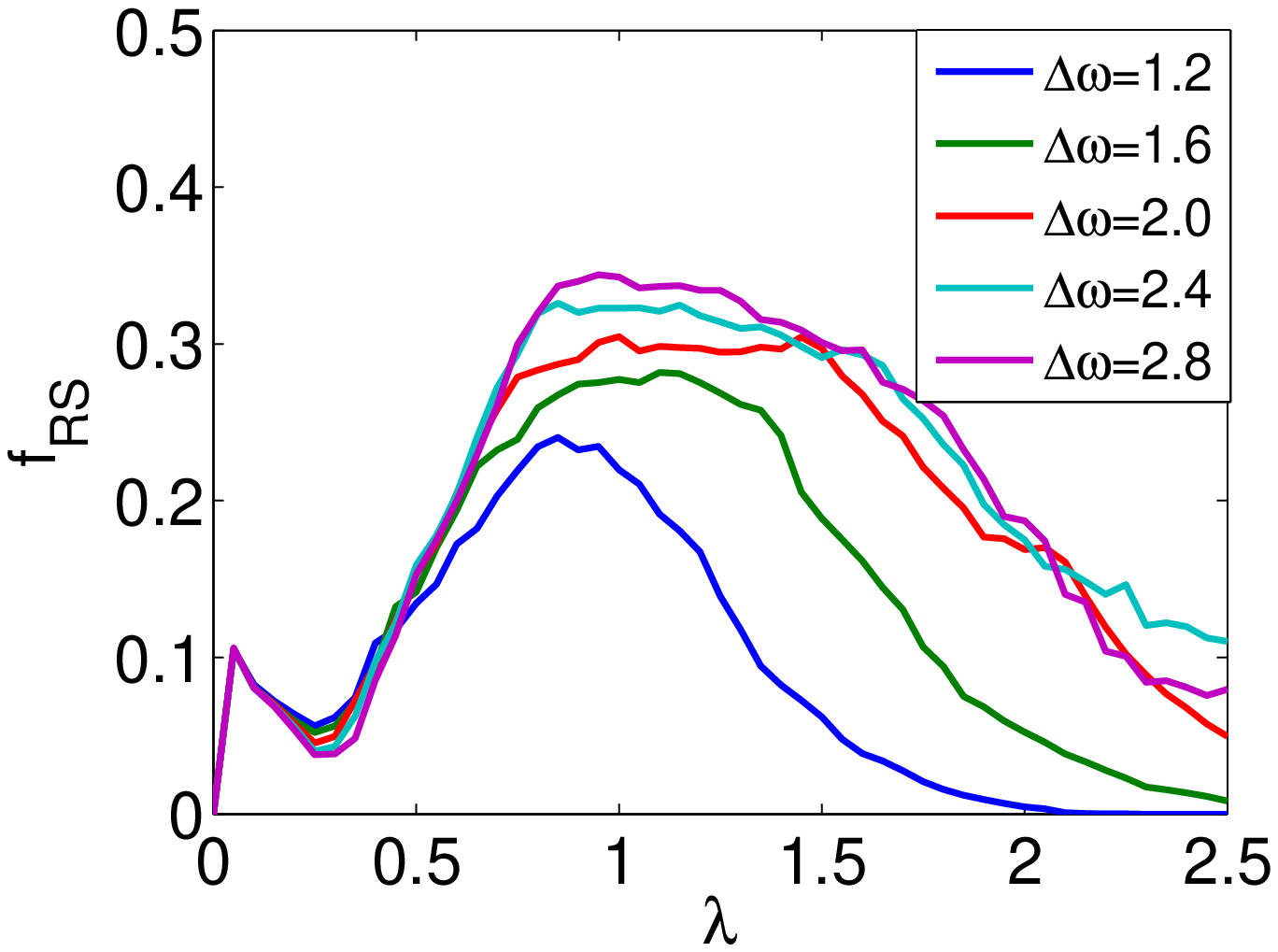}}
  \caption{(color online). Evolution of the fraction of RS links, $f_{RS}$, in SF (a) and ER (b) networks as a function of the coupling strength $\lambda$, and for different values of $\Delta \omega$. The remaining parameters are set as in Fig.~\ref{fig:diagrammi2D}. The fraction of RS links first increased as $\lambda$ is increased, with one (in ER networks) or two peaks (in SF networks) as observed for the evolution of $n_{RS}$, and then falls as networks recruit physical (instead of RS) links to get synchronized.
\label{fig:rB}}
\end{figure}

\begin{figure}
\centering
\subfigure[]{\includegraphics[height=.2\textheight]{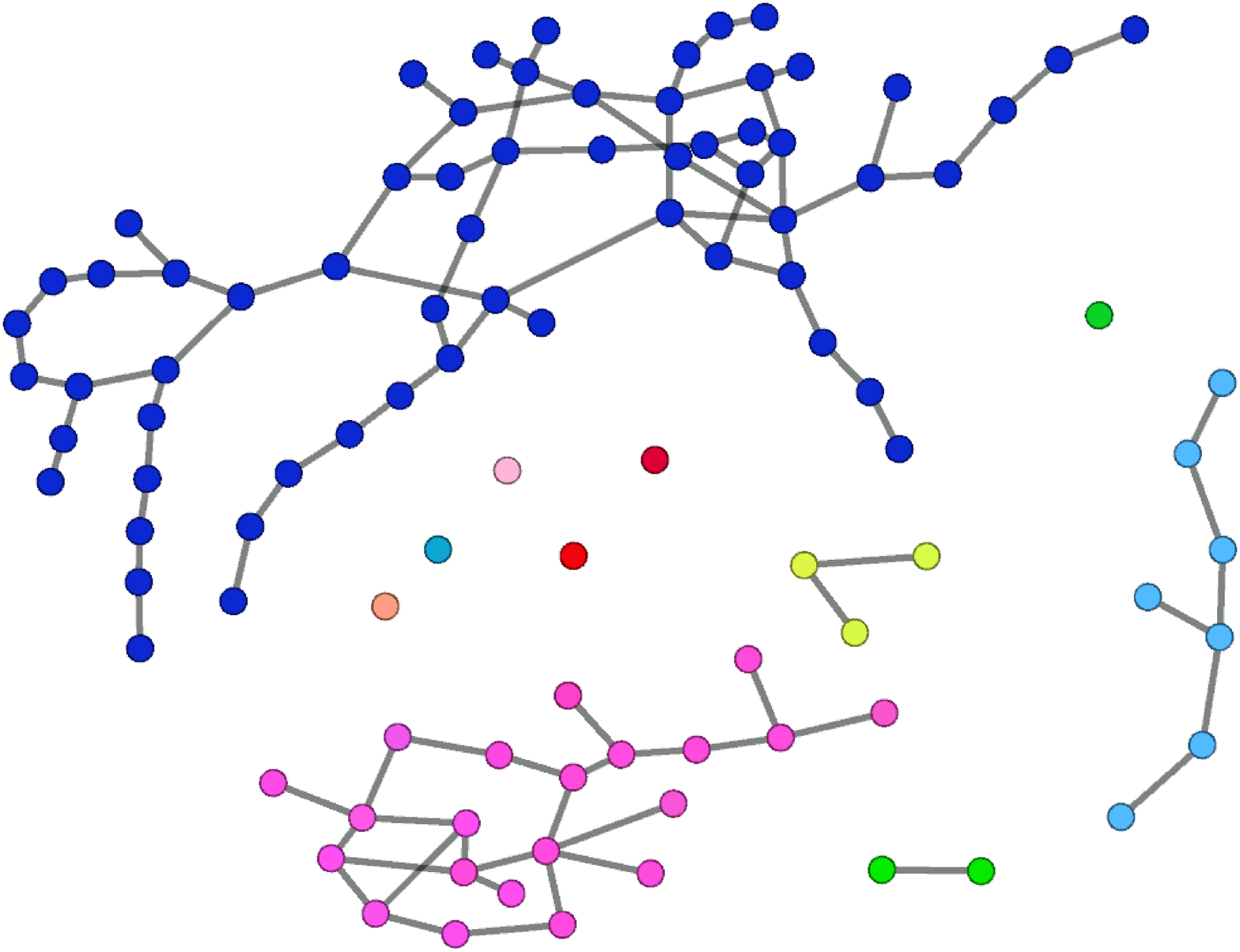}\label{fig:aggregationthroughRS_A}}
\subfigure[]{\includegraphics[height=.2\textheight]{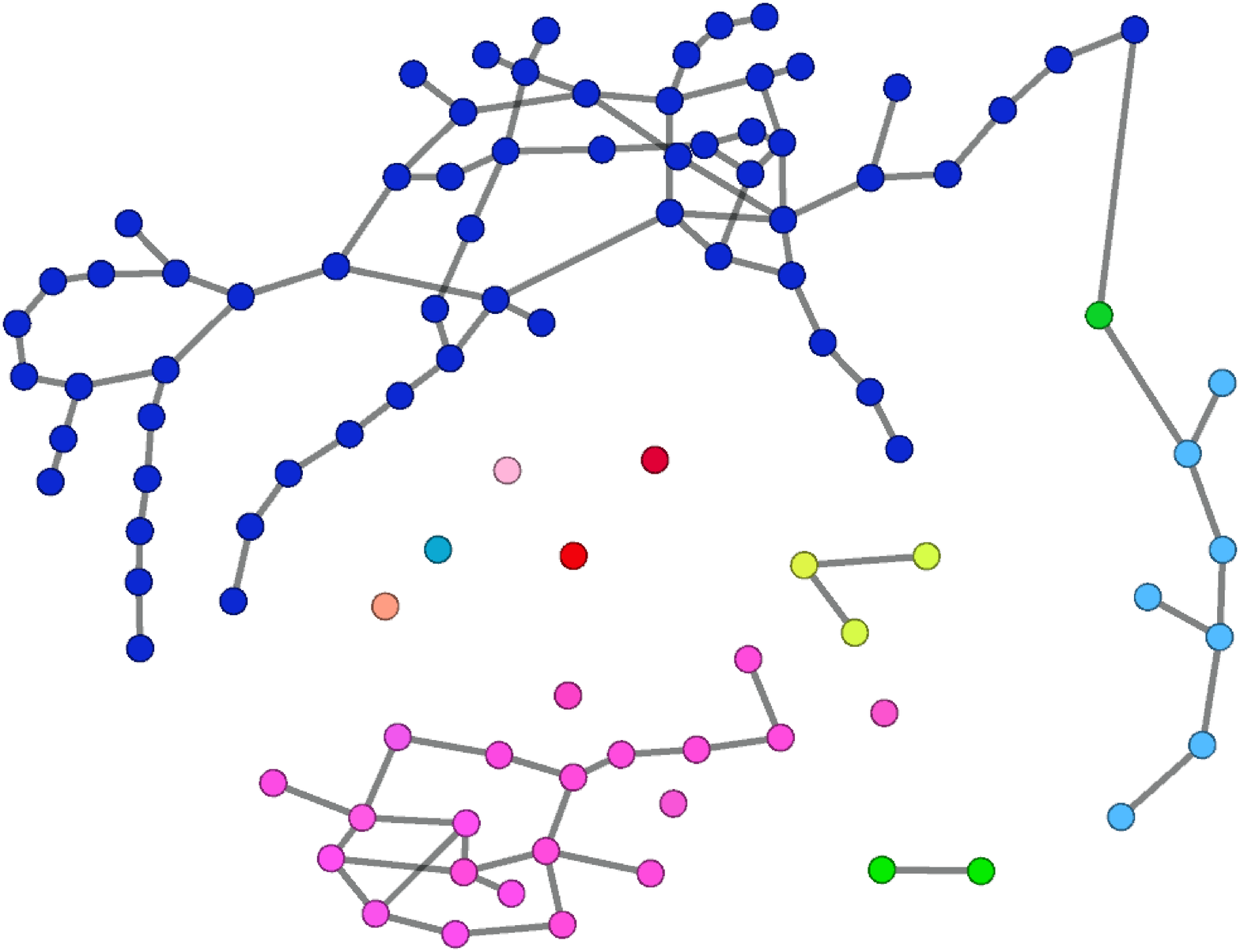}\label{fig:aggregationthroughRS_B}}\\
\subfigure[]{\includegraphics[height=.2\textheight]{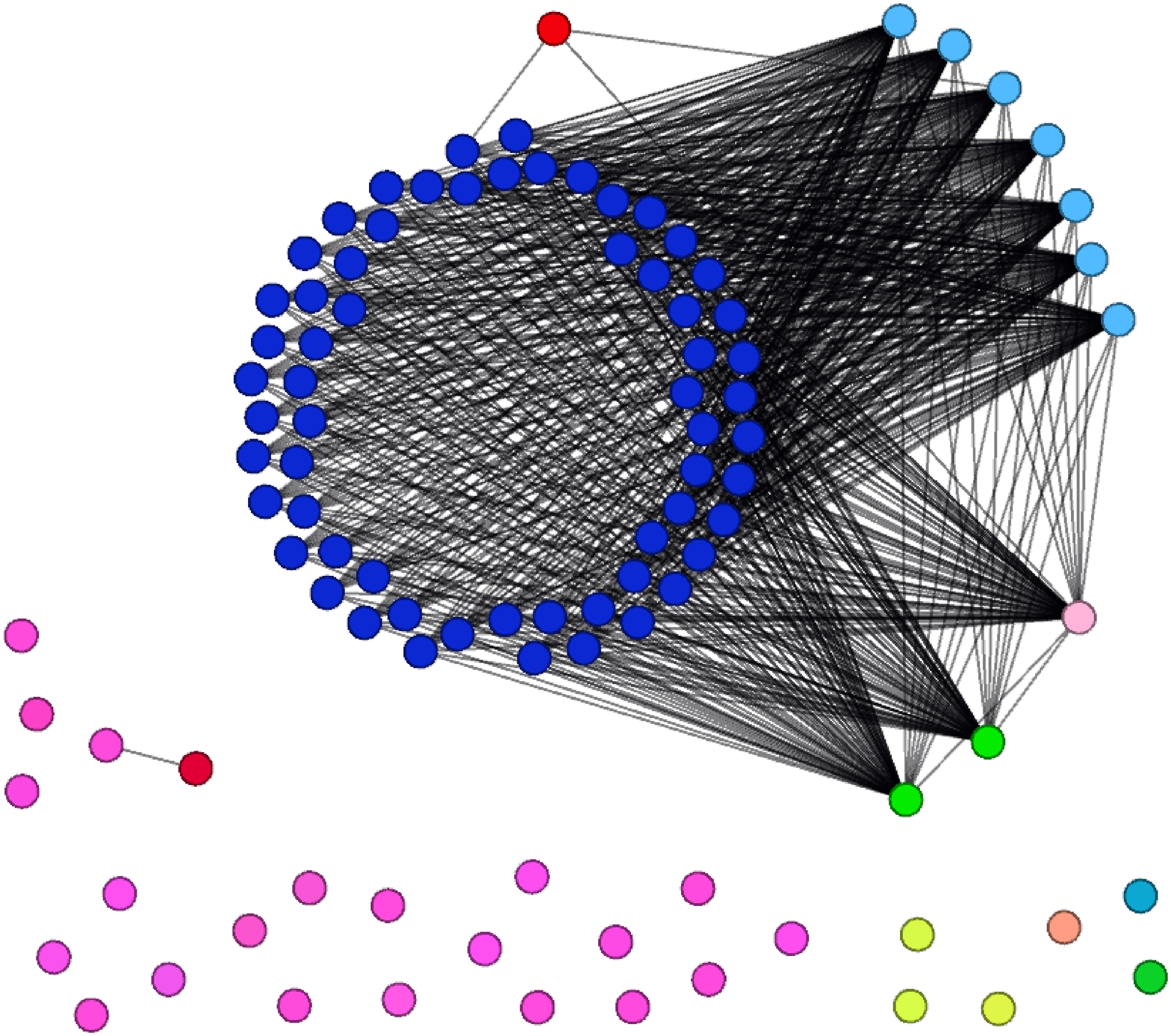}\label{fig:aggregationthroughRS_C}}
\subfigure[]{\includegraphics[height=.2\textheight]{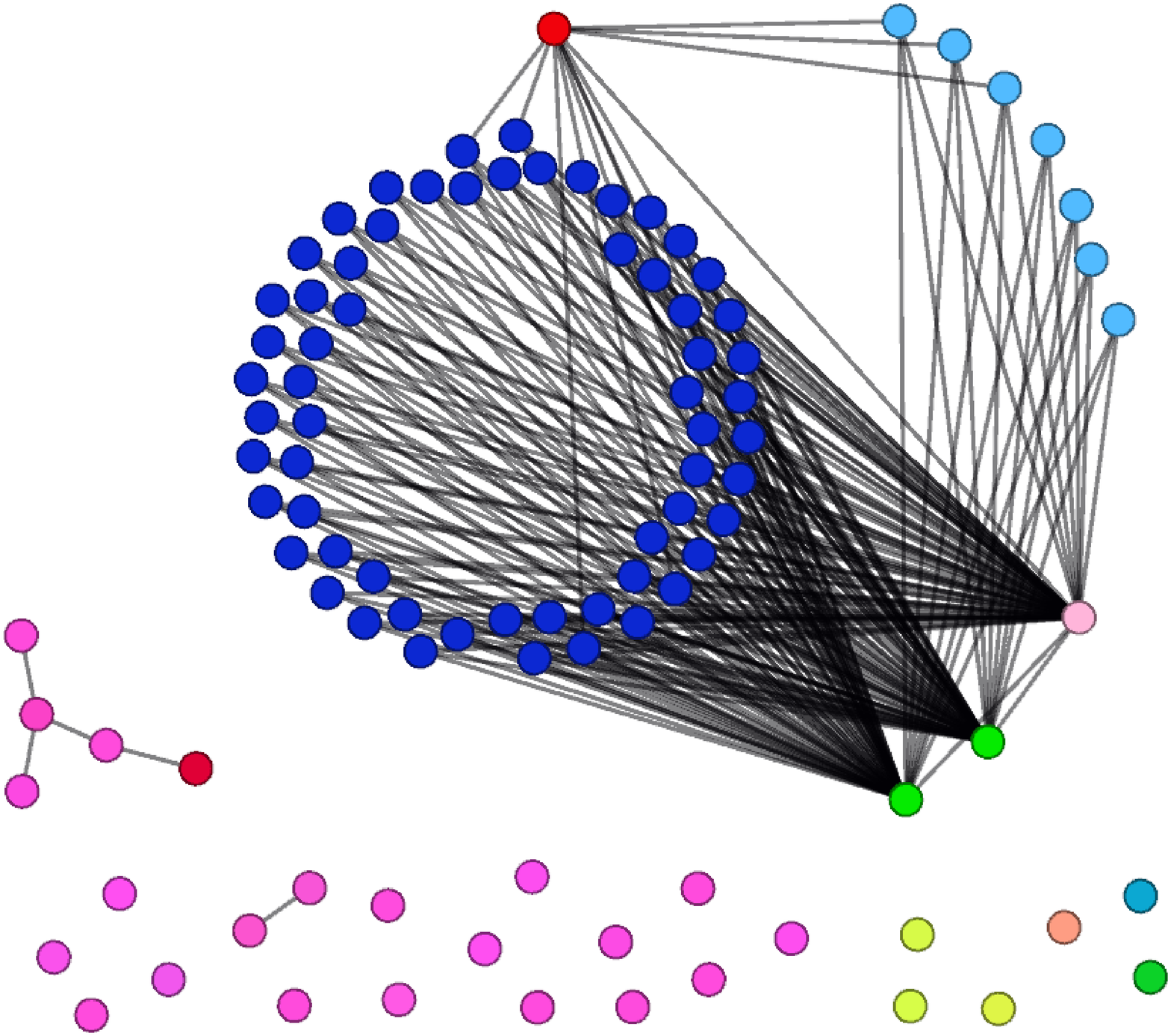}\label{fig:aggregationthroughRS_D}}
\caption{(color online). Evolution of components of physically (a)-(b) and remotely (c)-(d) synchronized nodes in an ER network with $\Delta \omega=2.8$, when $\lambda$ is increased by adiabatic continuation from $\lambda=1.65$ (a)-(c) to $\lambda=1.70$ (b)-(d). Nodes are colored according to the physically synchronized component they belong to when $\lambda=1.65$, {\em i.e.} in (a). The remaining parameters are the same as in Fig~\ref{fig:diagrammi2D}. At $\lambda=1.70$ two communities (the one with blue nodes and the one with cyan nodes), that were remotely synchronized at $\lambda=1.65$, fuse into a single one and, as a consequence, the RS links between the two communities existing for $\lambda=1.65$ disappear at $\lambda=1.70$.
\label{fig:aggregationthroughRS}}
\end{figure}

To gain more insight into the relation between the regime of remote synchronization and the onset of global synchrony we now consider the analysis of all synchronized pairs and its partition into those corresponding to remote synchronization and those for which a synchronized physical connection (either a direct link or a path of synchronized nodes) exists. To this aim, we define $\eta_{ij}=1$ if nodes $i$ and $j$ are connected either by a physical link or by a path of synchronized nodes and $\eta_{ij}=0$, otherwise. We then introduce the following quantities:
\begin{equation}
\label{eq:rP}
f_P=\frac{\sum_{i,j=1}^N{\eta_{ij} H(r_{ij}-\delta)}}{\sum_{i,j=1}^N{ H(r_{ij}-\delta)}}\,,
\end{equation}
\noindent and
\begin{equation}
\label{eq:rRS}
f_{RS}=\frac{\sum_{i,j=1}^N{(1-\eta_{ij}) H(r_{ij}-\delta) }}{\sum_{i,j=1}^N{ H(r_{ij}-\delta) }}\,,
\end{equation}
\noindent where $H(x)$ is the Heaviside function. Thus, $f_P$ and $f_{RS}$ represent the fraction of synchronized links due to a physical or remote connection, respectively. Obviously, as $f_{P}+f_{RS}=1$, it is enough to report the behavior of $f_{RS}$.

In Fig.~\ref{fig:rB} we show the evolution of $f_{RS}$ vs. $\lambda$ for several values of $\Delta \omega$. The presence of two peaks in the evolution of $f_{RS}$ in SF networks reveals a similar behavior to that found for $n_{RS}$. As $\Delta \omega$ increases, the percentage of RS links increases and the two peaks shift towards increasing values of $\lambda$. On the other hand, for ER networks the percentage of RS links is higher than in SF networks and $f_{RS}$ shows, as in the case of $n_{RS}$, a rise-and-fall trend. The fall in the number of RS links points out that the network is able to recruit physical links to get synchronized and thus those regions that appeared as RS become merged into a single component made of physically synchronized links.

To visualize the progressive substitution of RS links by physical ones in the path towards full synchronization we show in Fig.~\ref{fig:aggregationthroughRS} for an ER network (with $\Delta \omega=2.8$) snapshots of both remotely and physically synchronized links for two values of the coupling $\lambda$. In  Figs.~\ref{fig:aggregationthroughRS_A} and  \ref{fig:aggregationthroughRS_C} we plot two networks corresponding to physically and remotely synchronized links respectively when $\lambda=1.65$. In this case the network is divided into several clusters of physically synchronized nodes (the color of the nodes corresponds to the cluster of physically synchronized links they belong to) and some nodes of these clusters appear remotely synchronized with nodes belonging to different clusters [as shown in Fig.\ref{fig:aggregationthroughRS_C}]. When $\lambda$ is increased to $\lambda=1.70$, two of these clusters merge together [Fig.~\ref{fig:aggregationthroughRS_B}] through two physically synchronized links that connect each cluster to a new node synchronized to each of them. Thus, at $\lambda=1.70$ two communities, that were remotely synchronized at $\lambda=1.65$, fuse into a single one and, as a consequence, those RS links between the nodes of the two communities reported for $\lambda=1.65$ in Fig.~\ref{fig:aggregationthroughRS_C} disappear at $\lambda=1.70$ [Fig.~\ref{fig:aggregationthroughRS_D}]. We note that the choice of the threshold $\delta$ may impact on which nodes are assigned to which groups, although we have observed qualitatively similar results when the threshold is changed. In fact, the evolution of communities remains the same, although the value of $\lambda$ at which they merge may be slightly different.

\begin{figure*}[!t]
\centering
  \subfigure[]{\includegraphics[height=.16\textheight]{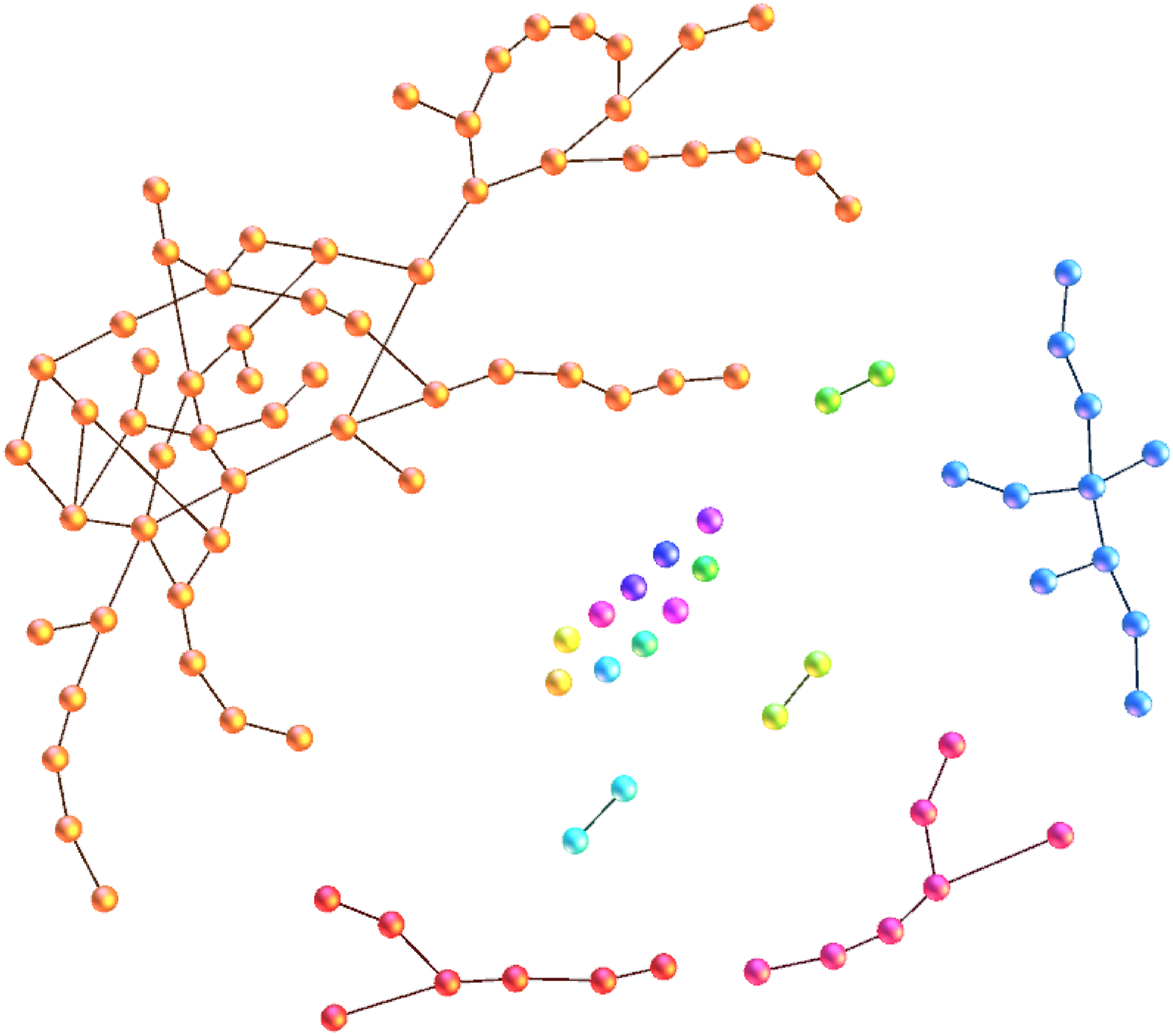}\label{fig:aggregationsecondexampleA}}
  \subfigure[]{\includegraphics[height=.16\textheight]{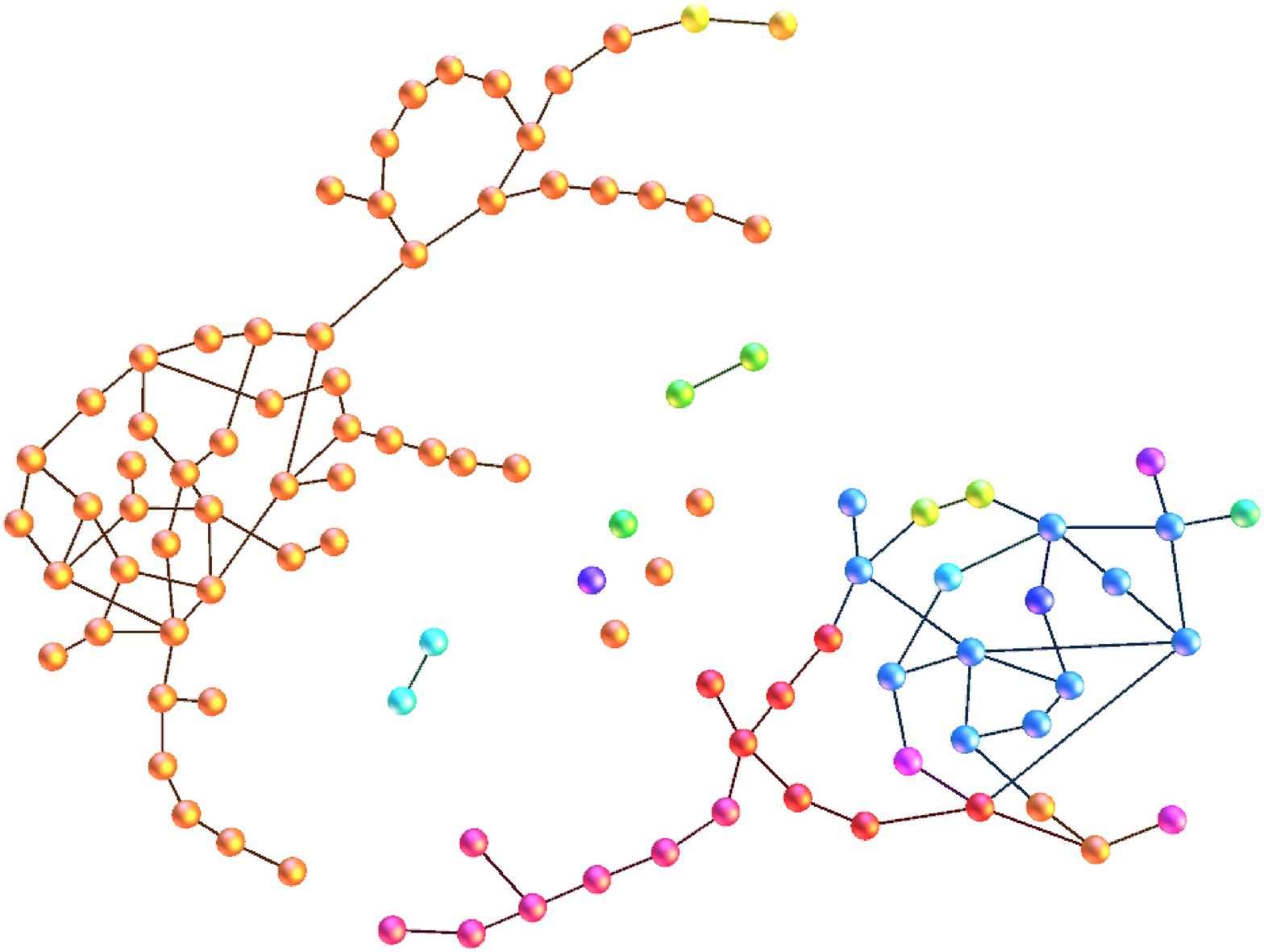}\label{fig:aggregationsecondexampleC}}
  \subfigure[]{\includegraphics[height=.16\textheight]{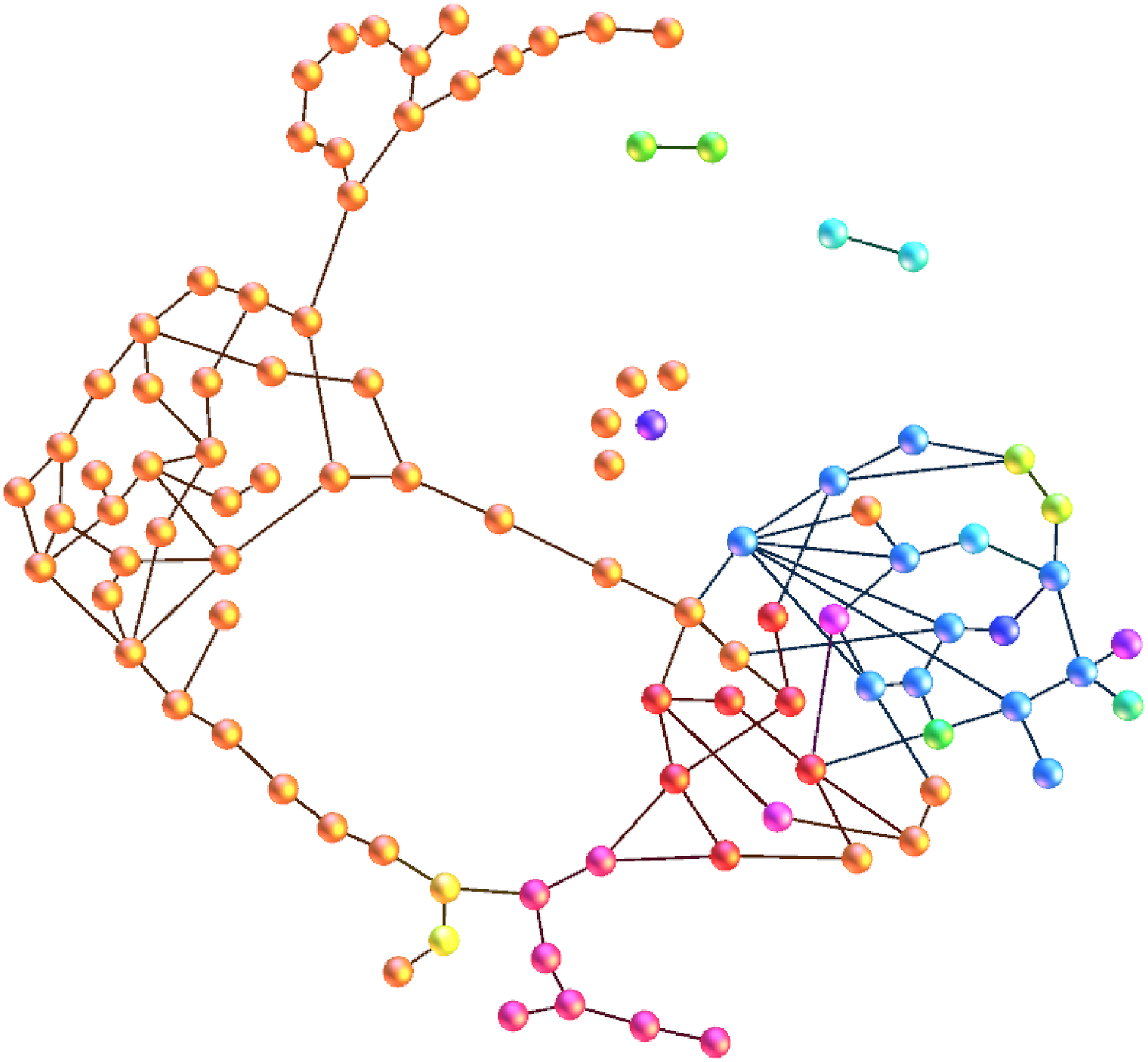}\label{fig:aggregationsecondexampleE}}
  \subfigure[]{\includegraphics[height=.16\textheight]{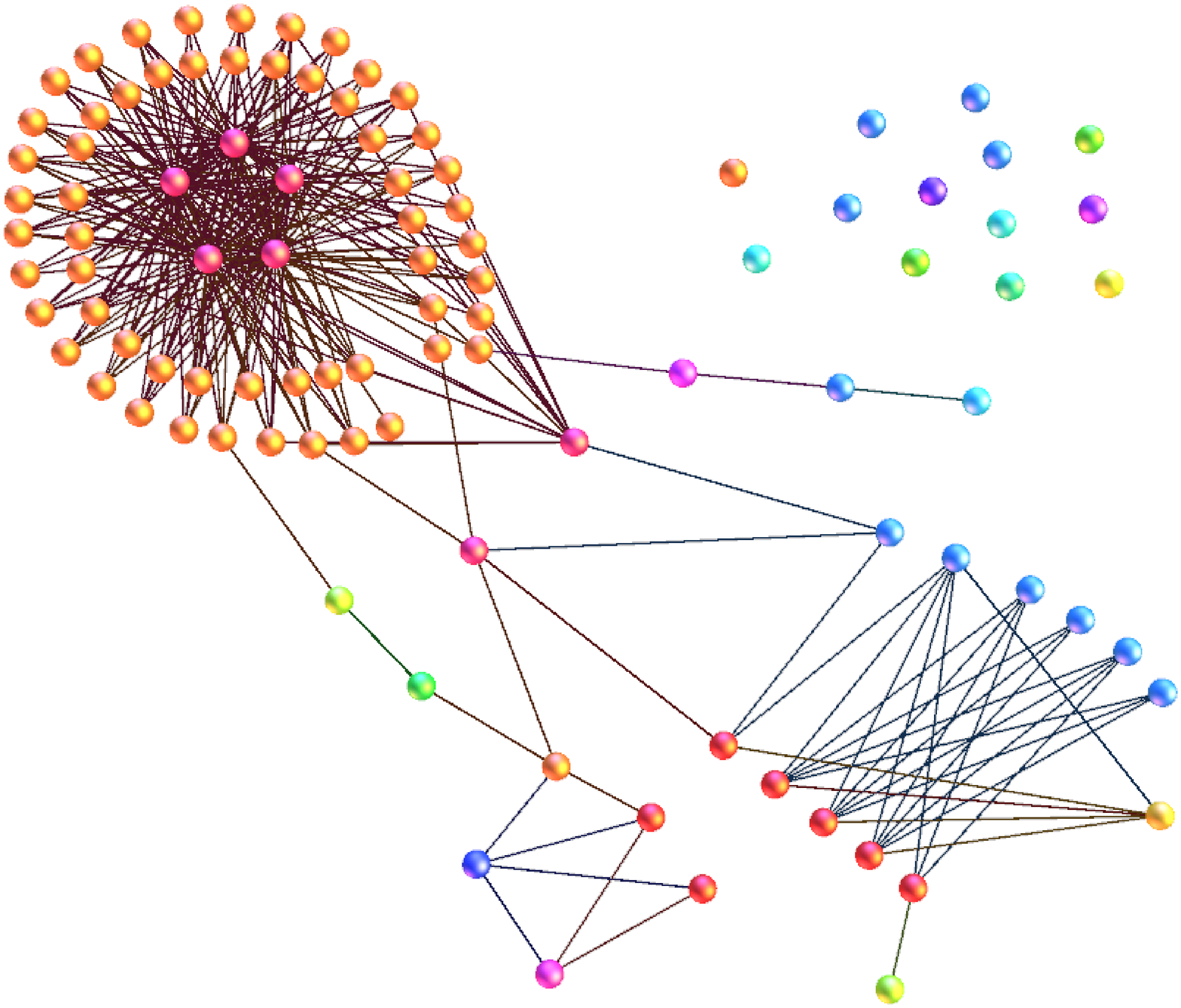}\label{fig:aggregationsecondexampleB}}
  \subfigure[]{\includegraphics[height=.16\textheight]{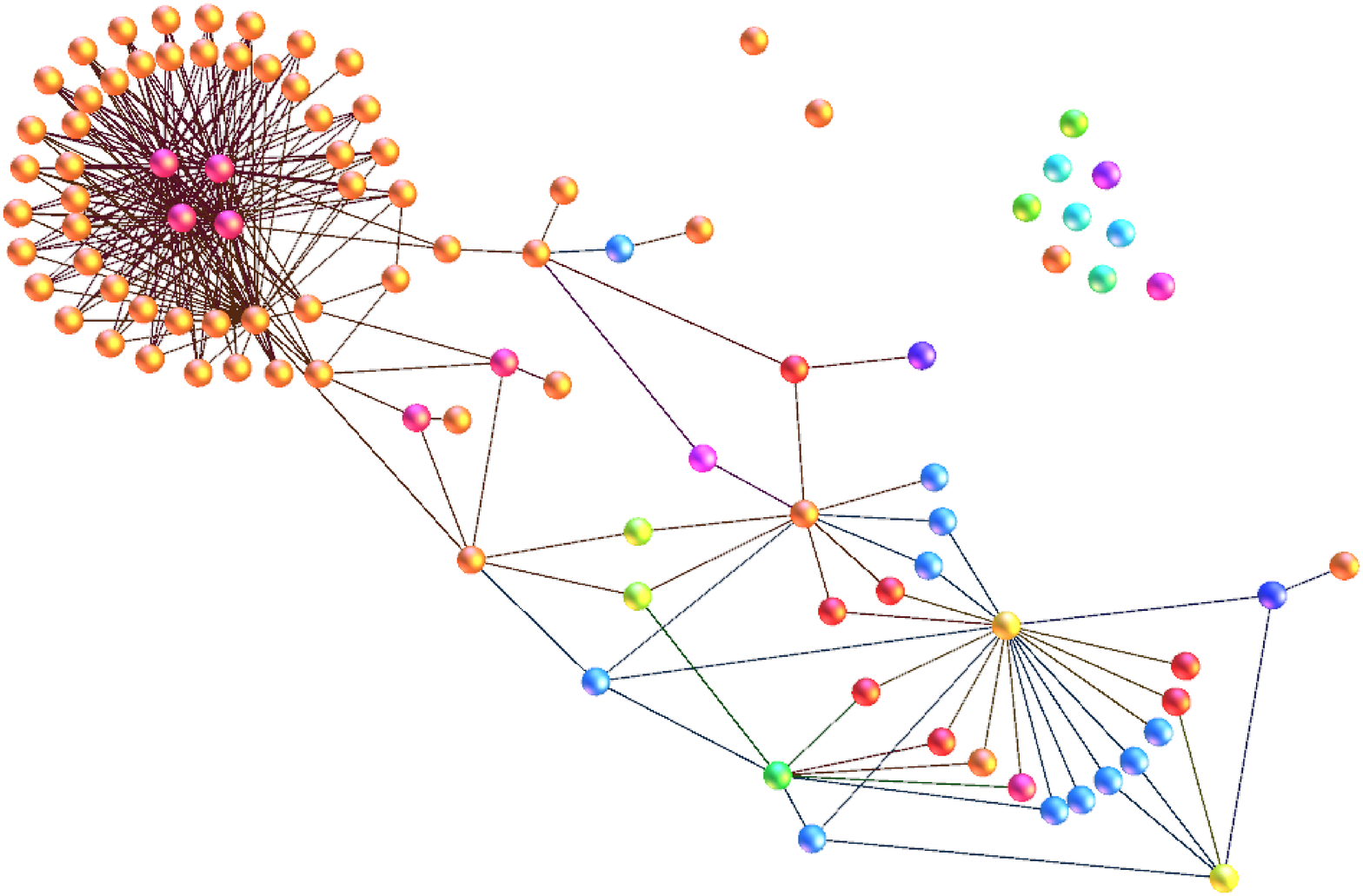}\label{fig:aggregationsecondexampleD}}
  \subfigure[]{\includegraphics[height=.16\textheight]{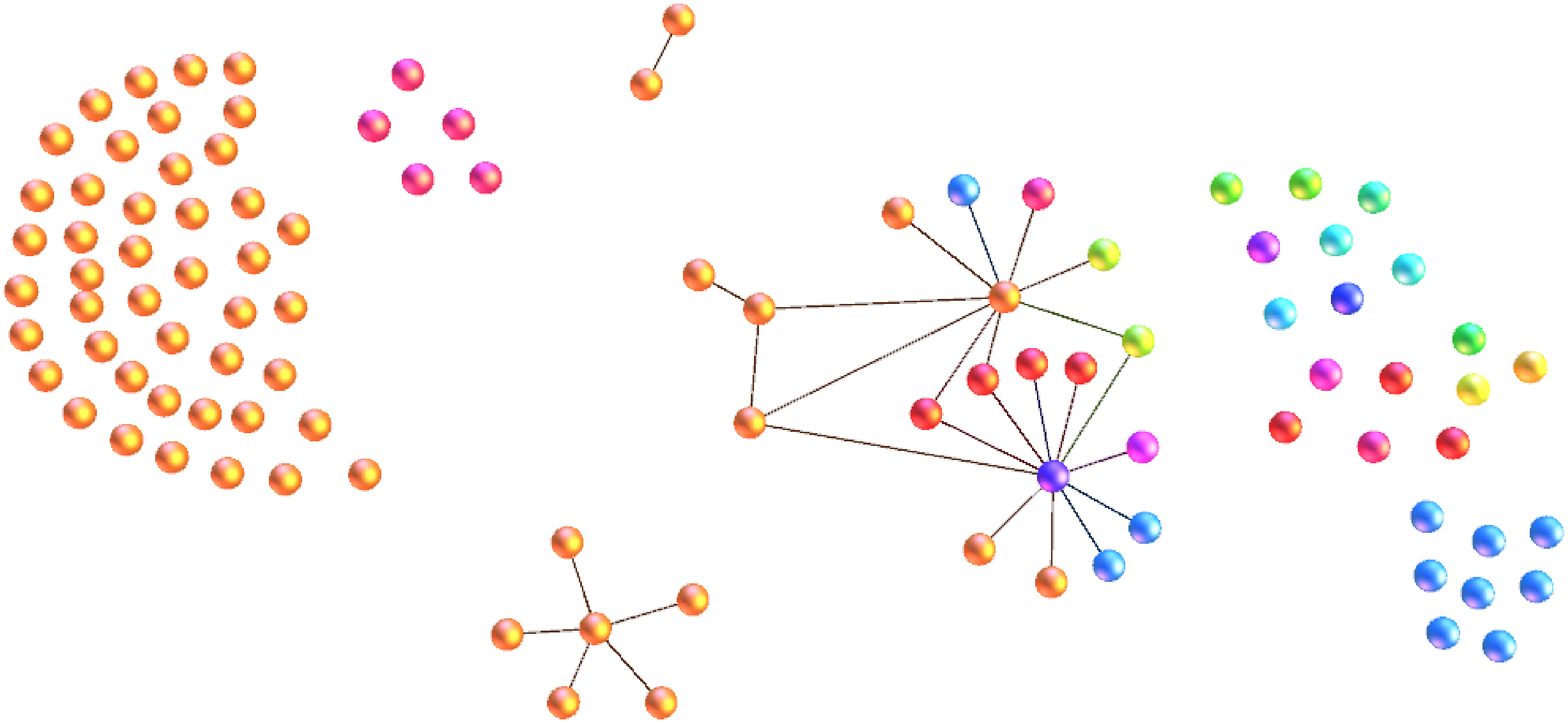}\label{fig:aggregationsecondexampleF}}
  \caption{Evolution of components of physically (a),(b), (c) and remotely (d),(e), (f) synchronized nodes in an ER network with $\Delta \omega=1.5$ and $\langle k\rangle=2$, when $\lambda$ is adiabatically increased. Nodes are colored according to the component to which they belong in (a) and, then, represented with the same colors in (b)-(f). The networks correspond to the following value of $\lambda$: (a)-(b) $\lambda=1.75$; (c)-(d) $\lambda=1.80$; (e)-(f) $\lambda=1.85$.
  For progressive increase of the coupling coefficient, first three of the four communities, existing at $\lambda=1.75$ and synchronized thanks to RS links, merge and, correspondingly, the RS links between these communities disappear, and, then, the fourth community (synchronized with the other three, already at  $\lambda=1.80$ thanks to RS links) aggregates to the previous ones.
  \label{fig:aggregationsecondexample}}
\end{figure*}

A further example of the merging of RS clusters is shown in Fig.~\ref{fig:aggregationsecondexample}. We consider an ER network with $\Delta \omega=1.5$ and $<k>=2$, when $\lambda$ is increased with continuation from $\lambda=1.75$ to $\lambda=1.85$. For $\lambda=1.75$ the network is divided into four main components of physically synchronized nodes plus some small communities and isolated nodes (Fig.~\ref{fig:aggregationsecondexampleA}). The analysis of the components of RS nodes (Fig.~\ref{fig:aggregationsecondexampleB}) reveals that there are RS links between the four communities. In fact, increasing the coupling to $\lambda=1.80$ three of these communities merge (Fig.~\ref{fig:aggregationsecondexampleC}) and, correspondingly, the RS links between these communities disappear (Fig.~\ref{fig:aggregationsecondexampleD}). Finally, a further increase of $\lambda$ ($\lambda=1.85$ in Fig.~\ref{fig:aggregationsecondexampleE}) leads to the aggregation of the fourth community (the bigger one) to the previous ones. Also in this case, almost all the RS links disappear (Fig.~\ref{fig:aggregationsecondexampleF}) and very few RS links are observed for $\lambda=1.85$.

\section{Conclusions}

In this paper we have provided measures to study remote synchronization in general complex networks. This phenomenon relies on the mutual synchronization of pairs of uncoupled nodes. Each remotely synchronized pair of nodes are thus physically connected through an intermediary node (not synchronized with them) or a sequence (path) of intermediary nodes. This is an important difference with another form of remote synchronization reported in \cite{vito}, where the analysis focused on the distribution of phase lags in a network of homogeneous oscillators (all oscillating at the same frequency) and a relationship between modules appearing in the network structure and the pattern of phase lags was revealed. The analysis we have presented reveals a stronger condition in that, according to our results, two RS nodes do not show any form of synchronization with intermediate nodes.

Although the original discovery of remote synchronization was restricted to a rather particular setup, a star graph, the analysis carried out in this paper, through the introduction of appropriate indicators, reveals that remote synchronization is common in general complex networks such as Erd\H{o}s-R\'enyi and Scale-free graphs of coupled oscillators having amplitude and phase as dynamical variables. The addition of amplitude as a dynamical variable, in contrast with the typical framework of networks of coupled phase-oscillators, provides the observation of remote synchronization and elucidates an important role played by it. In fact, we have found that remote synchronization constitutes a mechanism anticipating synchronization by physical links in networks with heterogeneous distribution of natural frequencies. Our results indicate that, in these networks, communities of nodes synchronized through RS links appear for values of coupling just lower than those allowing the merging of these communities through physical links. As synchronization is ubiquitous in natural and man-made systems, we suggest that this can be an important mechanism to explain the emergence of communities of synchronized nodes, not connected by physical links.

Our work suggests that remote synchronization is not significant for ensembles of phase oscillators, since its main underlying mechanism seems to be the modulation of the amplitude of intermediary nodes allowing information transfer between uncoupled pairs of nodes. In fact, when similar settings are applied to phase oscillators a different phenomenon is observed, namely that of \emph{explosive synchronization} \cite{explosive} in which the typical second-order synchronization transition transforms into a first-order one. In its turn, remote synchronization appears as a rather robust state prior the onset of global synchronization since for a wide range of coupling strengths almost all the nodes are remotely synchronized with, at least, another one while the level of global synchronization remains small. Thus, our results open the door for experimental observations of this novel state in which the existence of a synchronized pair cannot be associated to a given physical interaction through a single link of the network and highlight the important difference between the real ({\em i.e.} associated with physical links) and the functional ({\em i.e.} emerging from synchronization) connectivity of a network.


\begin{thebibliography}{abc99}
\bibitem{footnote} The value of $k^*$ is set so that the percentage of hubs is around the 20\% of the total number of nodes in the network.
\end{thebibliography}
\end{document}